\documentclass[12pt,letterpaper]{article}
\newif\ifscienceformat
\scienceformatfalse

\usepackage[T1]{fontenc}
\usepackage{amsmath,amssymb}
\usepackage{mathptmx}
\usepackage{graphicx}
\usepackage[letterpaper,margin=1in]{geometry}
\usepackage{scicite}
\usepackage{url}
\date{}
\renewenvironment{abstract}{\quotation}{\endquotation}

\makeatletter
\ifscienceformat\else
\fi
\renewcommand{\fnum@figure}{\textbf{Fig. \thefigure}}
\renewcommand{\fnum@table}{\textbf{Table \thetable}}
\newcommand{\bibnumber}[1]{\@ifundefined{b@#1}{??}{\csname b@#1\endcsname}}
\makeatother
\newcommand{\chisq}{\chi^2}
\newcommand{\pqm}{\textsc{PQMass}}
\newcommand{\Nbin}{N_\mathrm{bin}}
\newcommand{\Nobs}{N_\mathrm{obs}}
\newcommand{\Nref}{N_\mathrm{ref}}

\graphicspath{{figures/}{figures_final/paper/}{figures_final/other/}}

\newcommand{\scititle}{Beyond $\chi^2$:\\Ensemble null tests for reliable scientific discovery}
\title{\bfseries\scititle}
\author{\parbox{0.97\textwidth}{\centering
Adrian E. Bayer$^{1,2,3,4,5,6\ast}$ and David N. Spergel$^{1}$\\[0.8em]
\small$^{1}$Center for Computational Astrophysics, Flatiron Institute, New York, NY, USA\\[0.3em]
\small$^{2}$Department of Astrophysical Sciences, Princeton University, Princeton, NJ, USA\\[0.3em]
\small$^{3}$The NSF AI Institute for Artificial Intelligence and Fundamental Interactions, Cambridge, MA, USA\\[0.3em]
\small$^{4}$Laboratory for Nuclear Science, Massachusetts Institute of Technology, Cambridge, MA, USA\\[0.3em]
\small$^{5}$Center for Astrophysics $|$ Harvard \& Smithsonian,
Cambridge, MA, USA\\[0.3em]
\small$^{6}$Perimeter Institute for Theoretical Physics, Waterloo, Ontario, Canada\\[0.3em]
\small$^\ast$Corresponding author. Email: abayer@mit.edu.
}}

\newcommand{\SchematicFigure}{%
\begin{figure}[!t]
\centering
\includegraphics[width=\textwidth]{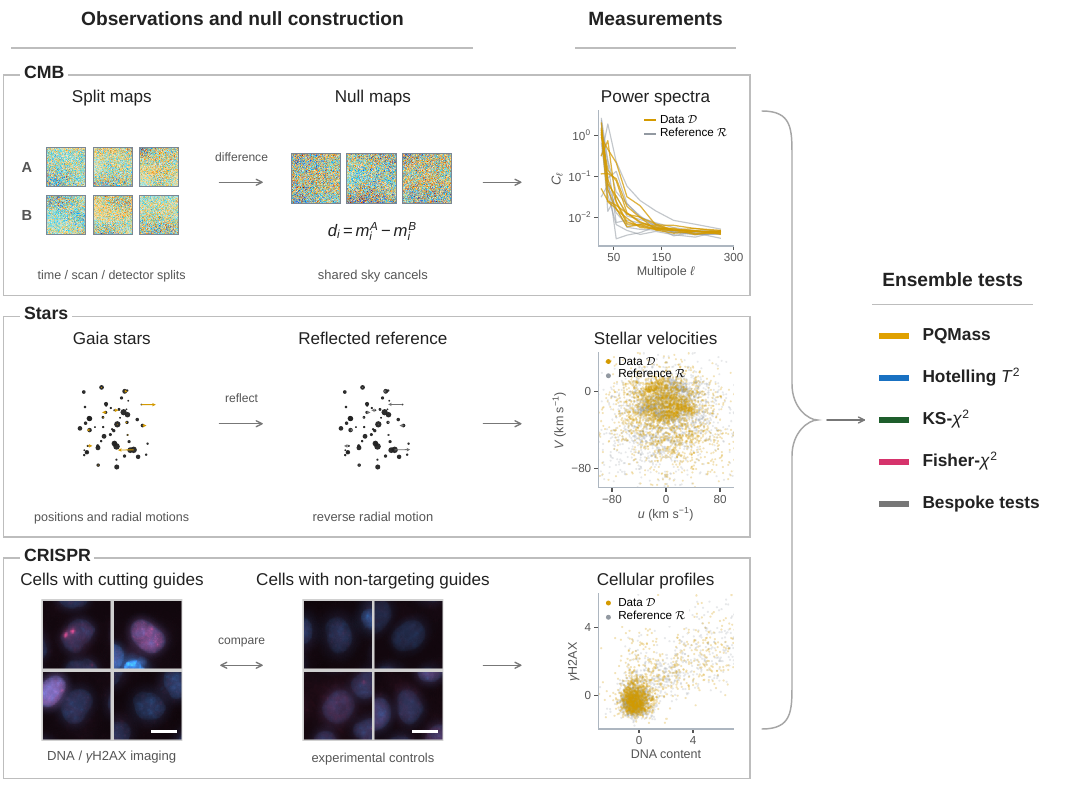}
\caption{\textbf{From observations to ensemble null tests.}
The three rows follow the same workflow: construct a reference, describe each
observation by a set of measurements, and compare the resulting ensembles.
CMB split maps are differenced to cancel the shared sky signal, with simulated
null maps providing the reference. Gaia stellar velocities are compared with
velocities reflected in the radial direction, and CRISPR cells carrying cutting guides
with non-targeting controls. Each map, star, or cell contributes measurements
of power across angular scales, velocity components, or cellular features.
Ensemble tests compare means, covariance-weighted distances, or full joint
distributions. The latter retain structure that simpler summaries can hide.
Bespoke tests target selected signatures within each application.}
\label{fig:schematic}
\end{figure}
}

\newcommand{\ControlledExamplesFigure}{%
\ifscienceformat
\begin{figure}[p]
\else
\begin{figure}[!t]
\fi
\centering
\includegraphics[width=\textwidth]{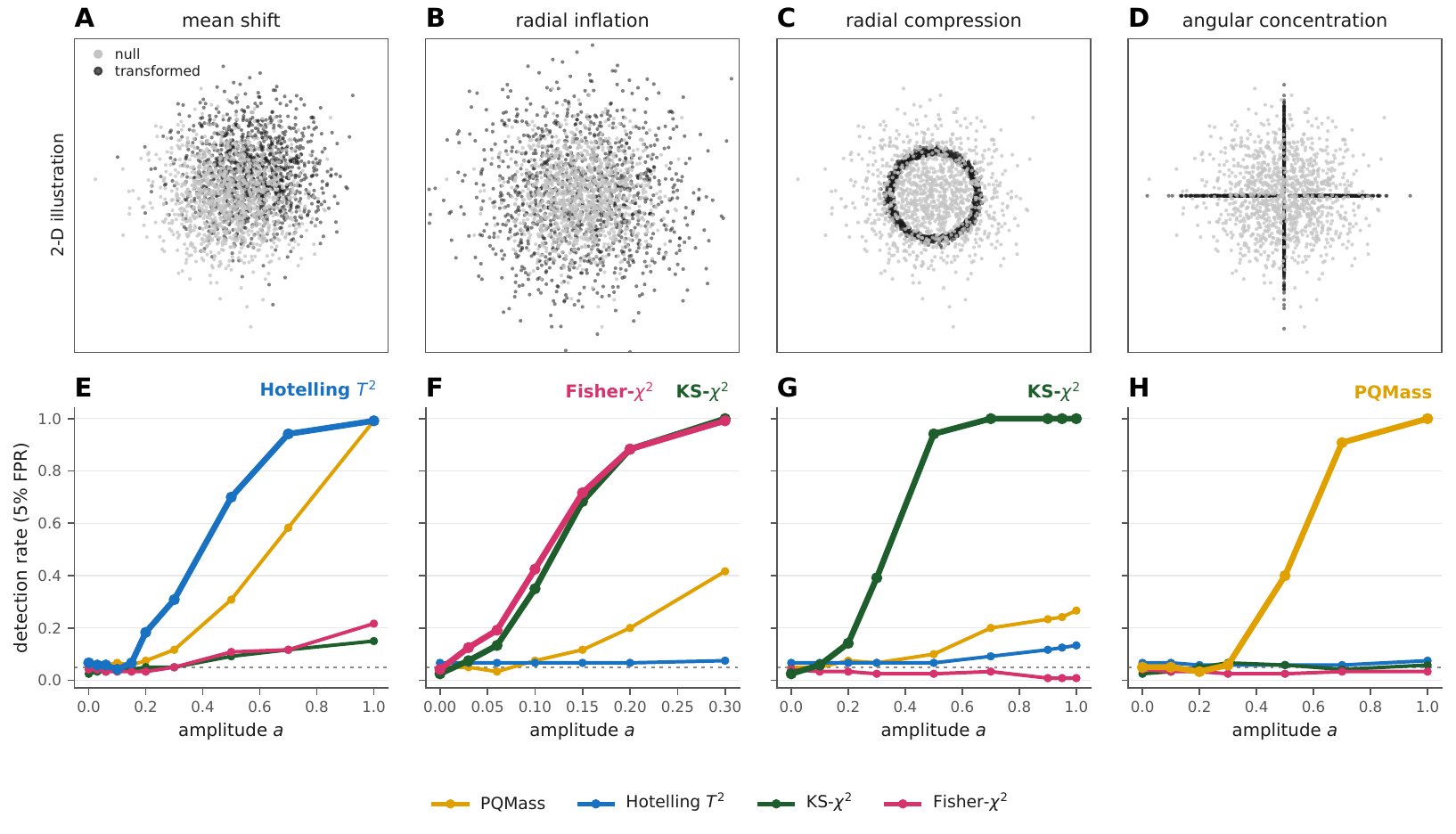}
\caption{\textbf{Complementary sensitivities of ensemble null tests.}
(\textbf{A}--\textbf{D}) Two-dimensional illustrations show a cloud of
measurements shifting, expanding, concentrating into a shell, or gathering
along preferred directions (gray, null; black, transformed).
(\textbf{E}--\textbf{H}) Detection rates for analogous changes in nine dimensions.
Bold curves highlight the most sensitive tests in each example; dashed lines mark the nominal 5\%
false-positive rate. Different tests excel for different departures.
\pqm{} remains sensitive to all four, including angular structure hidden
from the mean and radial summaries.}
\label{fig:rung-illustration}
\label{fig:ladder}
\end{figure}
}

\newcommand{\CoadditionComparisonFigure}{%
\ifscienceformat
\begin{figure}[p]
\else
\begin{figure}[!t]
\fi
\centering
\includegraphics[width=\textwidth]{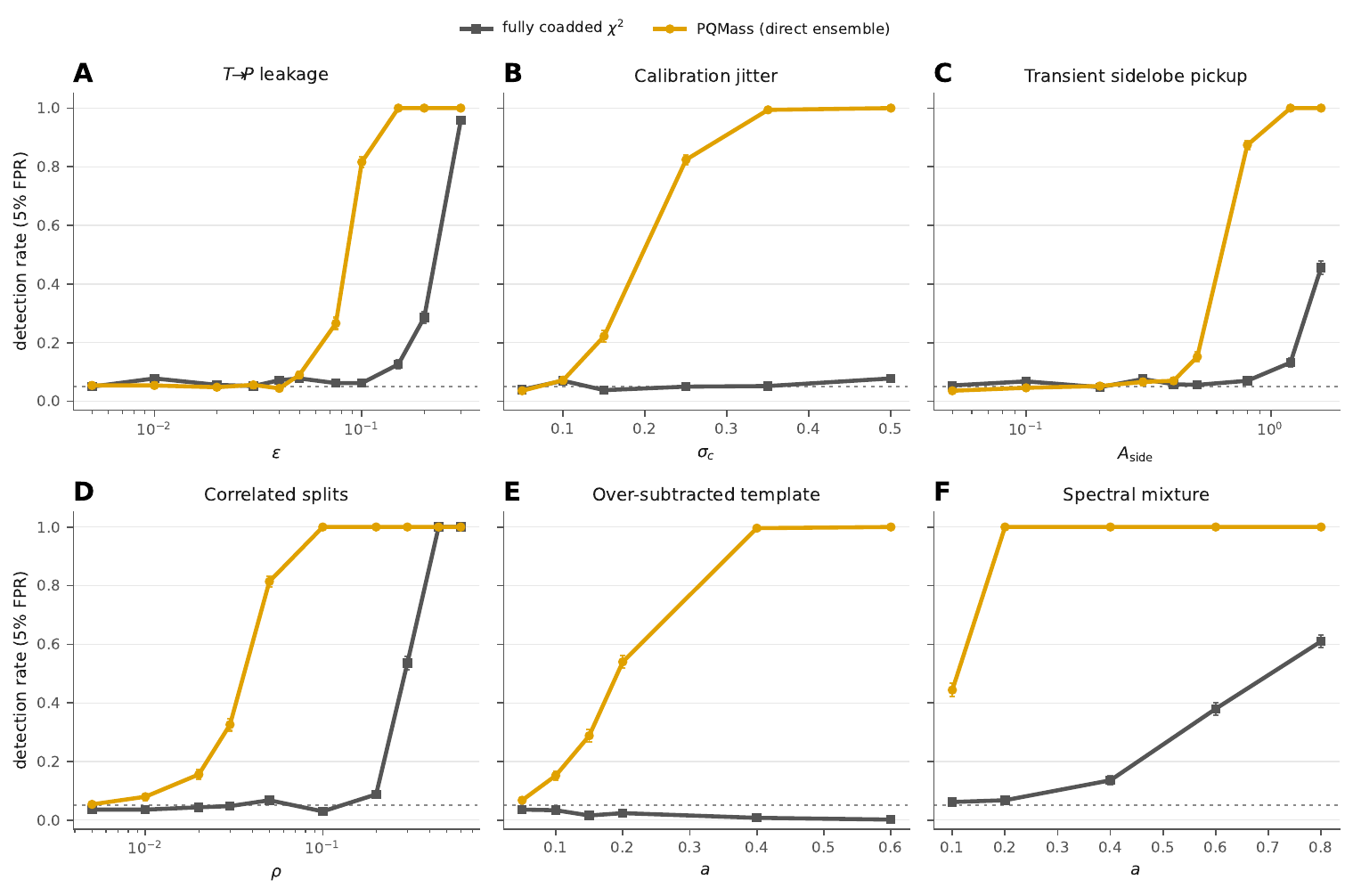}
\caption{\textbf{Averaging can hide detectable contamination.}
The CMB maps are analyzed as an ensemble of $100$ null spectra
with \pqm{} (orange), or coadded into a single null and tested with $\chisq$
(gray). Curves show how often each approach detects contamination at a nominal
5\% false-positive rate; error bars show binomial uncertainty across independent
mock experiments. The ensemble comparison detects weaker contamination in
the same observations, retaining variation that coaddition suppresses.}
\label{fig:twosplit}
\end{figure}
}

\newcommand{\PQMassMethodFigure}{%
\ifscienceformat
\begin{figure}[p]
\else
\begin{figure}[!htbp]
\fi
\centering
\includegraphics[width=\textwidth]{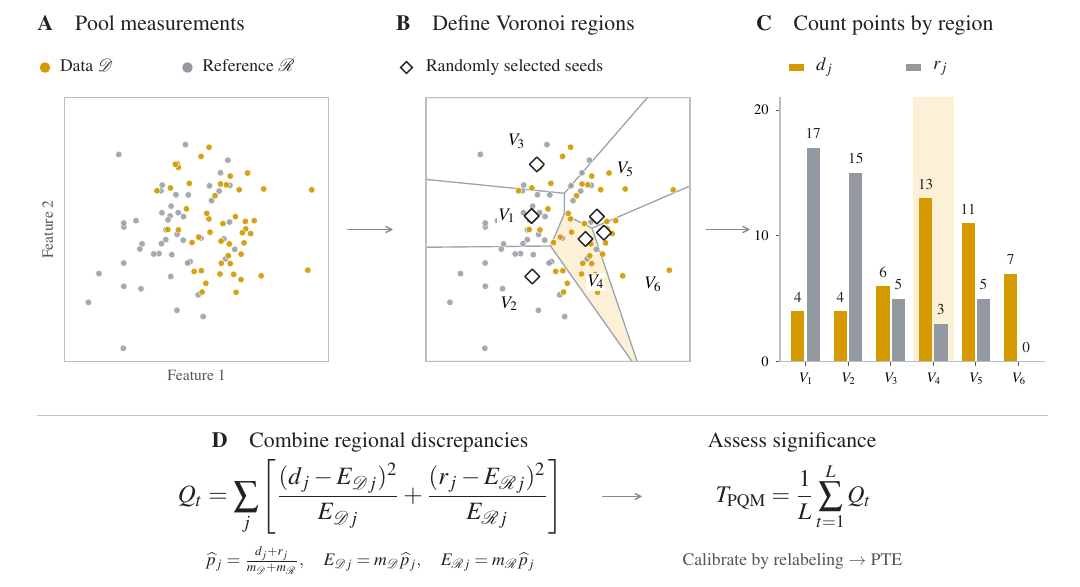}
\caption{\textbf{How \pqm{} compares ensembles through regional counts.}
This two-dimensional illustration uses fixed synthetic points.
(\textbf{A}) Data (gold) and reference (gray) measurements. (\textbf{B}) Six randomly selected seeds (open diamonds) define Voronoi
regions. Each measurement point is assigned to its nearest seed. 
(\textbf{C}) Counts of the measurements in each region. The shaded region and bar pair both identify $V_4$.
(\textbf{D}) The Pearson statistic compares the observed counts with
their pooled expectations, combining regional discrepancies across the multivariate measurement space. The analysis averages this statistic
over $L$ tessellations and calibrates it using permuted ensemble labels. The six regions shown here illustrate the construction;
the analysis uses the setup described in section~\ref{app:tests}.}
\label{fig:pqmass-method}
\end{figure}
}

\newcommand{\BaselineNoiseFigure}{%
\ifscienceformat
\begin{figure}[p]
\else
\begin{figure}[!htbp]
\fi
\centering
\includegraphics[width=\textwidth]{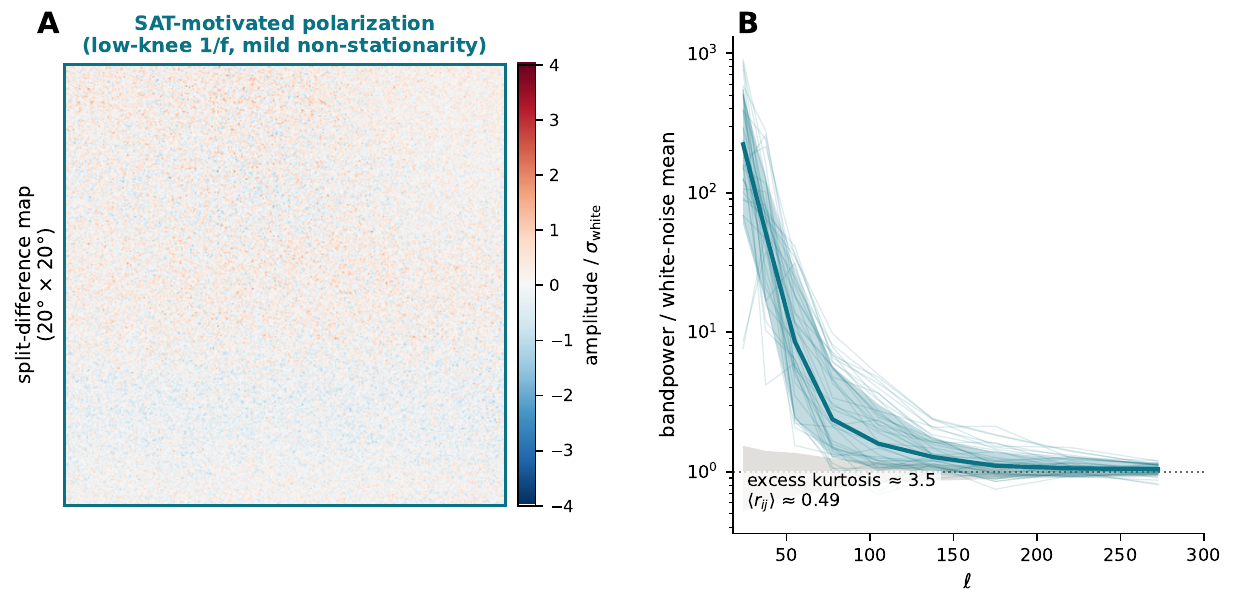}
\caption{\textbf{The SAT-motivated polarization null.} Left: one
observation-level split-difference map $d_i$. Right: the corresponding
binned auto-spectra for one ensemble (thin curves), with their median
and central 68\% interval, shown relative to the white-noise mean; the
gray band is the corresponding central interval of the white-noise
null, and the annotations give the pooled (per-bin standardized)
bandpower excess kurtosis and
mean off-diagonal bin--bin correlation.
Residual $1/f$ power is concentrated at low multipoles, while depth
variation and scan-aligned striping introduce mild non-stationarity.
(\textbf{A}) Split-difference map. (\textbf{B}) Bandpower ensemble.}
\label{fig:noisetypes}
\end{figure}
}

\newcommand{\SignaturesDetectionFigure}{%
\ifscienceformat
\begin{figure}[p]
\else
\begin{figure}[!tbp]
\fi
\centering
\includegraphics[width=\textwidth]{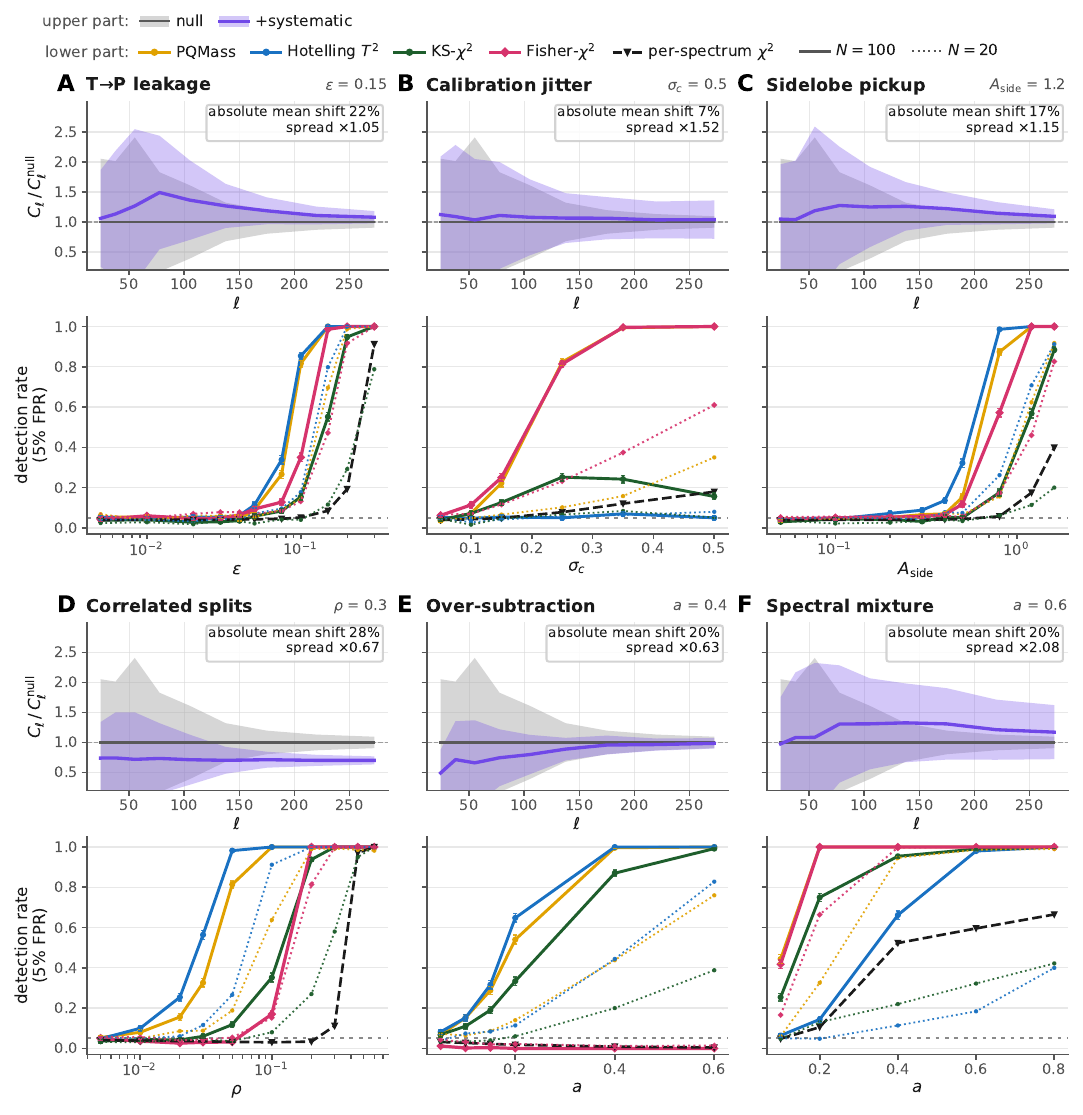}
\caption{\textbf{Distributional signatures and detection of six CMB systematics.}
Upper plots show how each contaminant changes the mean null spectrum and
its variation between observations (gray, baseline; purple, contaminated;
bands, $\pm1\sigma$). Values are normalized by the baseline mean; annotations
summarize changes in the mean and scatter.
Lower plots show how often each test detects the contamination at a nominal
5\% false-positive level, using ensembles of $N=100$ (solid) or $N=20$ (dotted)
spectra. \pqm{} (orange), Hotelling's $T^2$ (blue), KS-$\chisq$ (green), and
Fisher-$\chisq$ (pink). The per-spectrum $\chisq$ curve (dashed black) gives
the mean fraction of individual spectra flagged by a $\chi^2$ test for $N=100$.
Error bars show uncertainty across independent mock experiments.
Different statistics excel for different systematics, while \pqm{} retains
sensitivity to all six without specifying their signatures in advance.}
\label{fig:signatures}
\end{figure}
}

\newcommand{\CoverageFigure}{%
\ifscienceformat
\begin{figure}[p]
\else
\begin{figure}[!htbp]
\fi
\centering
\includegraphics[width=0.85\textwidth]{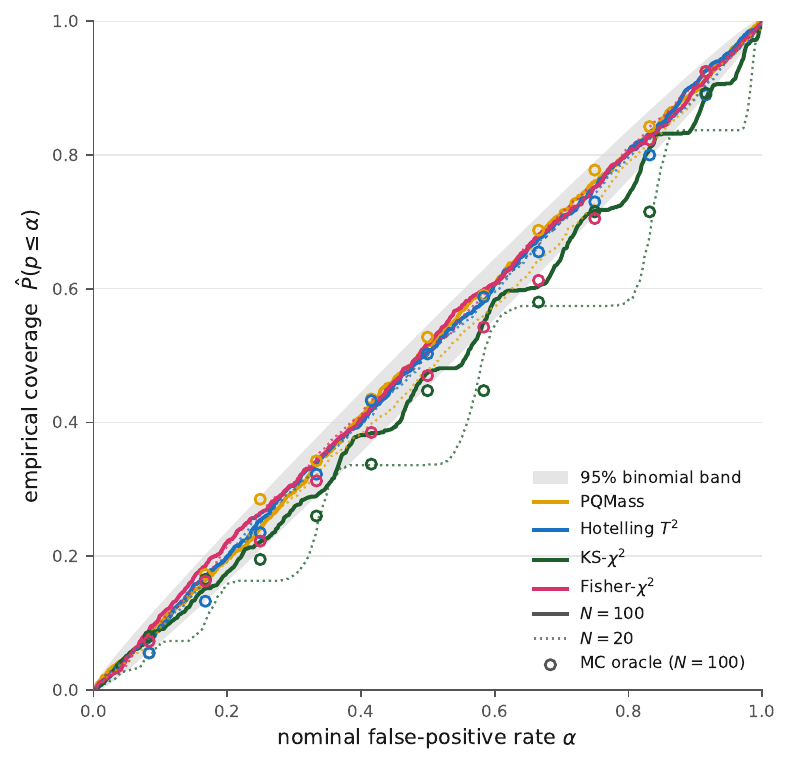}
\caption{\textbf{Achieved rejection probability $\widehat P(p\leq\alpha)$ versus nominal
false-positive rate $\alpha$ on uncontaminated experiments at $N=100$ and
$N=20$.} The diagonal denotes exact calibration; curves below it are
conservative. Solid curves use the production calibration. Open circles show an
oracle calibration from an external bank of independent clean ensemble pairs
(400 evaluations). Bands show pointwise binomial uncertainty.
All tests are well calibrated, except KS-($\chi^2$) with $N=20$, which is mildly conservative.}
\label{fig:coverage-pp}
\end{figure}
}

\newcommand{\PermutationRanksFigure}{%
\ifscienceformat
\begin{figure}[p]
\else
\begin{figure}[!htbp]
\fi
\centering
\includegraphics[width=\textwidth]{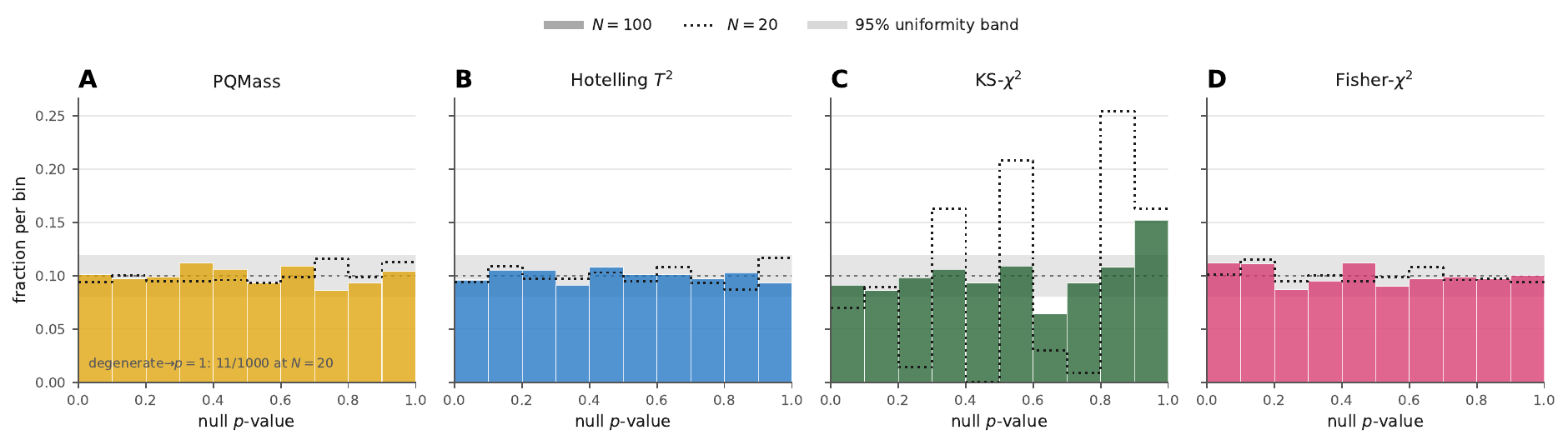}
\caption{\textbf{Global-test PTE distributions on the same uncontaminated experiments.} The horizontal band gives the pointwise expectation for continuous uniform
PTEs; discrete permutation and rank statistics can instead be conservative.
The PTE distributions are broadly consistent with null uniformity, implying all tests are well calibrated, except KS-($\chi^2$) with $N=20$, which is mildly conservative.
Panels show (\textbf{A}) \pqm{}, (\textbf{B}) Hotelling's $T^2$, (\textbf{C}) KS-$\chisq$, and (\textbf{D}) Fisher-combined $\chisq$.}
\label{fig:coverage-rank}
\end{figure}
}

\newcommand{\CalibrationComparisonFigure}{%
\ifscienceformat
\begin{figure}[p]
\else
\begin{figure}[!htbp]
\fi
\centering
\includegraphics[width=\textwidth]{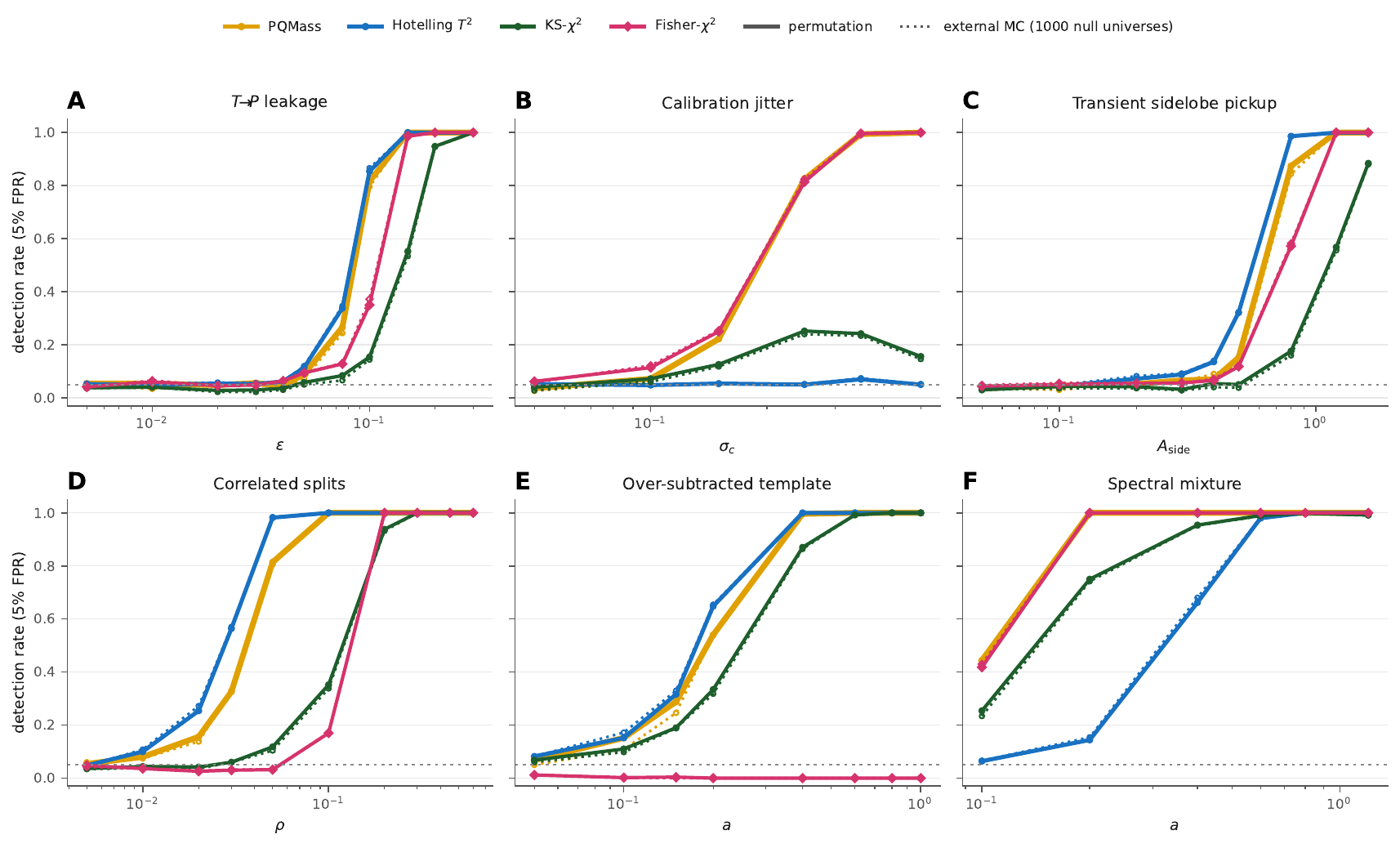}
\caption{\textbf{The $N=100$ detection curves of main-text Figure~\ref{fig:signatures} under both calibration routes:
solid, the permutation threshold computed within each experiment; dotted, the
same statistic ranked against a fixed bank of $1000$ null universes with
$\mathcal D$ and $\mathcal R$ both null, under the same add-one rank convention.} The routes differ by $-0.002$ on average, rms $0.009$, over the $200$ (scenario,
amplitude, test) combinations shown; $26$ differ by more than twice their
discordance error, the largest by $0.044$.
Panels show (\textbf{A}) temperature-to-polarization leakage, (\textbf{B}) calibration jitter, (\textbf{C}) transient sidelobe pickup, (\textbf{D}) correlated splits, (\textbf{E}) an over-subtracted template, and (\textbf{F}) a spectral mixture.}
\label{fig:calibgallery}
\end{figure}
}

\newcommand{\MultivariateComparisonFigure}{%
\ifscienceformat
\begin{figure}[p]
\else
\begin{figure}[!htbp]
\fi
\centering
\includegraphics[width=\textwidth]{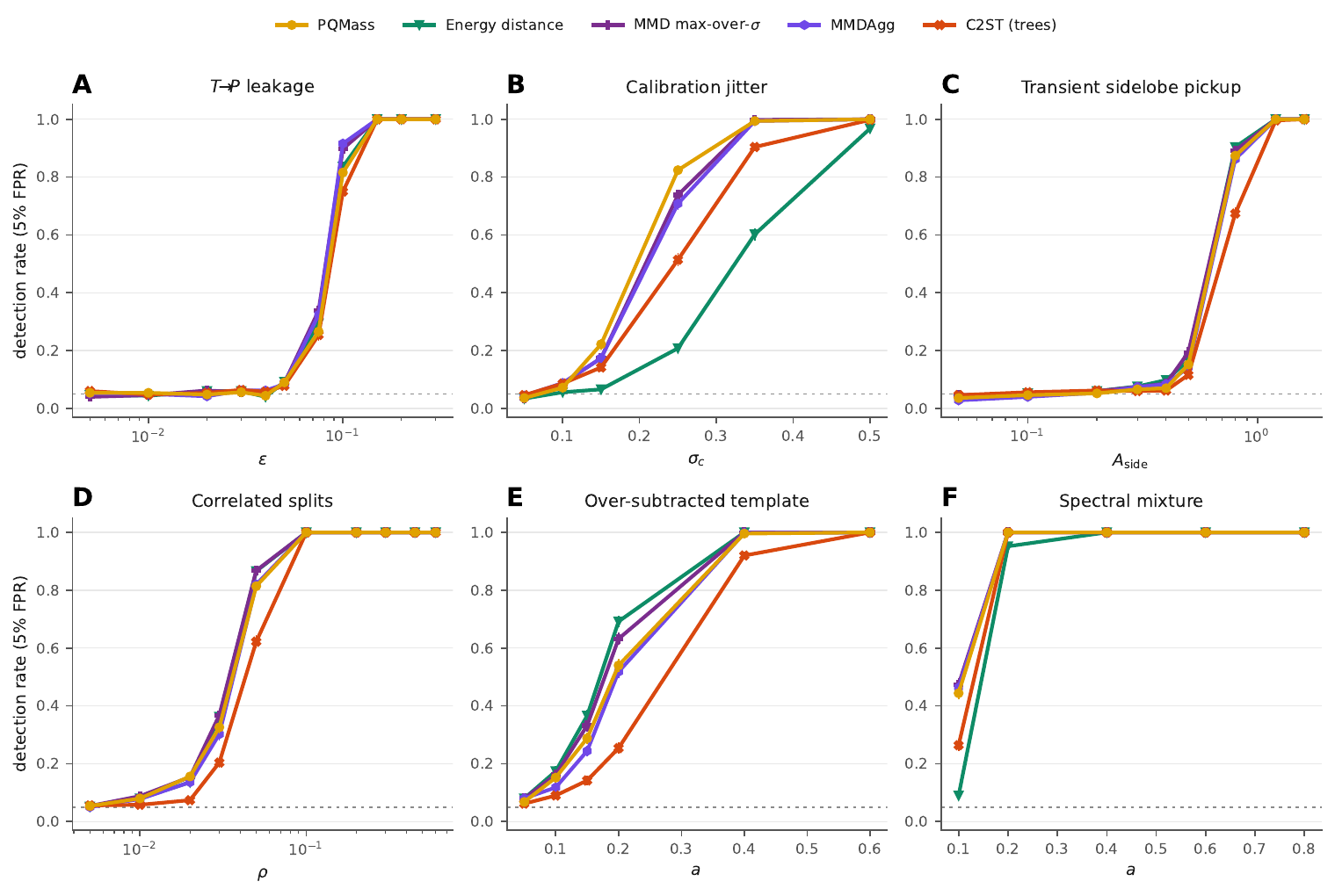}
\caption{\textbf{Sensitivity across multivariate ensemble tests.}
Each panel shows experiment-level rejection rates for one
CMB-motivated systematic, using $N=100$ measurements in each
ensemble and $\Nbin=9$.
Rates are estimated from 500 independent experiments per
contamination amplitude at a nominal 5\% rejection level,
indicated by the dashed horizontal line.
Colors identify methods; error bars indicate binomial uncertainty.
\pqm{}, energy distance, MMD max-over-$\sigma$ (max-MMD),
and C2ST trees use $B=500$ label permutations.
MMDAgg uses its published aggregation and calibration procedure,
with $B_1=B_2=2000$ and $B_3=50$
(section~\ref{app:multivariate-comparison}).
Panels show (\textbf{A}) temperature-to-polarization leakage,
(\textbf{B}) calibration jitter, (\textbf{C}) transient sidelobe
pickup, (\textbf{D}) correlated splits, (\textbf{E}) an
over-subtracted template, and (\textbf{F}) a spectral mixture.}
\label{fig:multivariate-comparison}
\end{figure}
}

\newcommand{\CoaddDepthFigure}{%
\ifscienceformat
\begin{figure}[p]
\else
\begin{figure}[!htbp]
\fi
\centering
\includegraphics[width=0.88\textwidth]{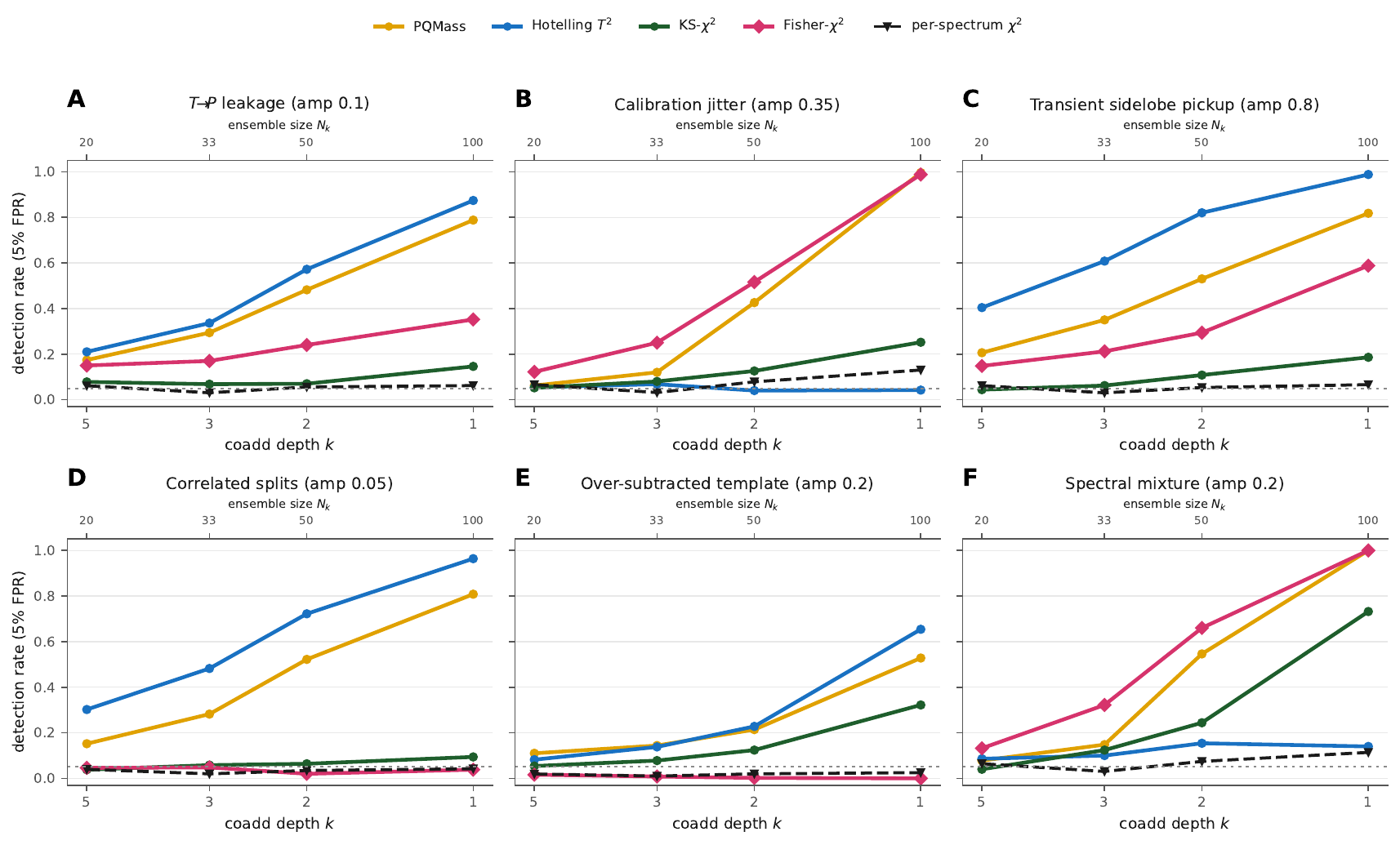}
\caption{\textbf{Response versus coadd depth $k$ at fixed total data (depth increases to
the left).} The top axis gives $N_k=\lfloor\Nobs/k\rfloor$. Curves show global
rejection rates for the ensemble tests and the mean per-spectrum $\chisq$ flag
fraction at one representative amplitude of each systematic.
Panels show (\textbf{A}) temperature-to-polarization leakage, (\textbf{B}) calibration jitter, (\textbf{C}) transient sidelobe pickup, (\textbf{D}) correlated splits, (\textbf{E}) an over-subtracted template, and (\textbf{F}) a spectral mixture.}
\label{fig:apjl-coadding-tradeoff}
\end{figure}
}

\newcommand{\EnsembleSizeFigure}{%
\ifscienceformat
\begin{figure}[p]
\else
\begin{figure}[!htbp]
\fi
\centering
\includegraphics[width=0.88\textwidth]{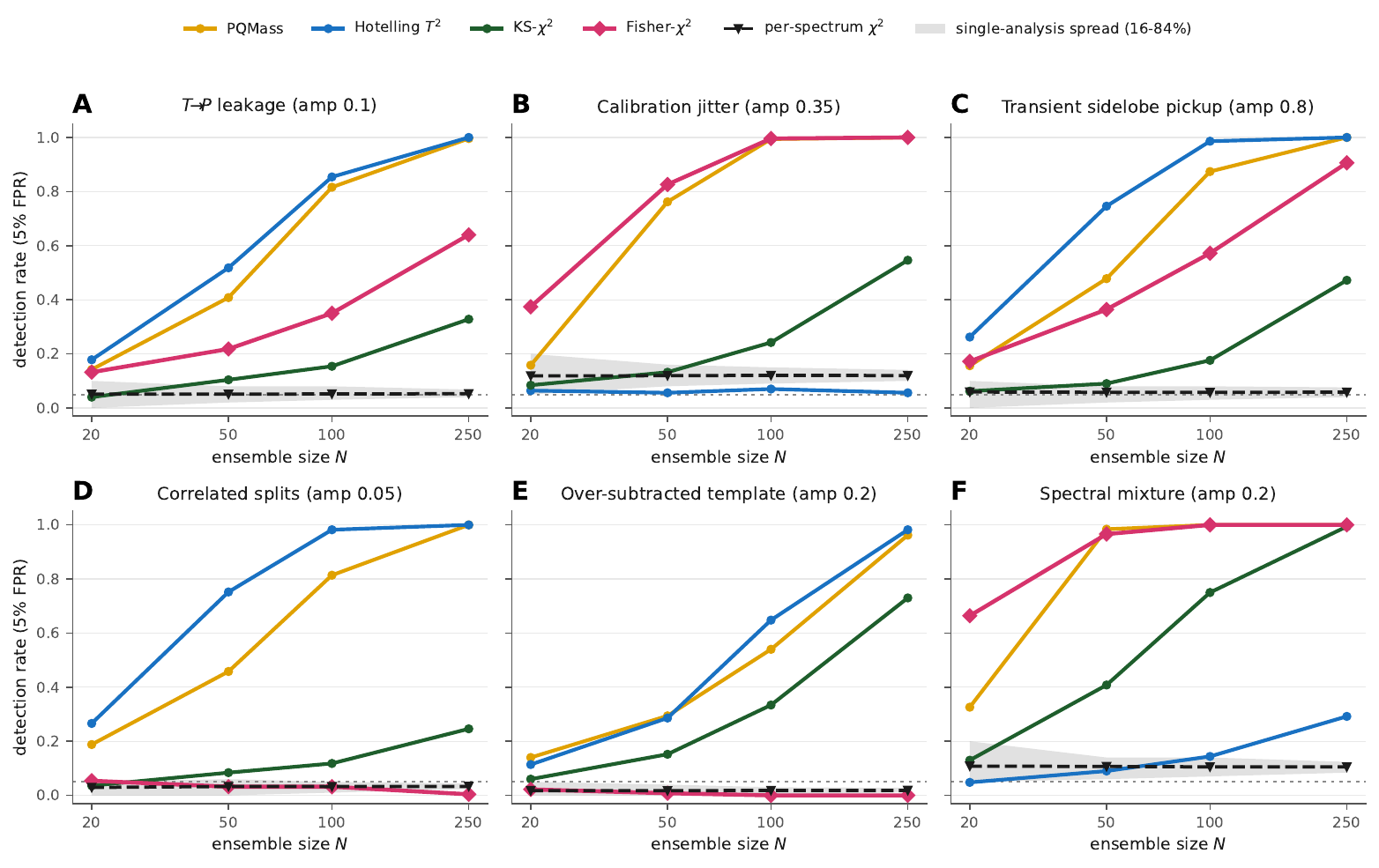}
\caption{\textbf{Response versus ensemble size $N$ at fixed per-spectrum noise and
representative contamination amplitudes.} Curves show global rejection rates for
the ensemble tests and the mean per-spectrum $\chisq$ flag fraction; the band
around the latter shows its central 68\% range across mock experiments.
Panels show (\textbf{A}) temperature-to-polarization leakage, (\textbf{B}) calibration jitter, (\textbf{C}) transient sidelobe pickup, (\textbf{D}) correlated splits, (\textbf{E}) an over-subtracted template, and (\textbf{F}) a spectral mixture.}
\label{fig:msweep}
\end{figure}
}

\newcommand{\ReferenceSizeFigure}{%
\ifscienceformat
\begin{figure}[p]
\else
\begin{figure}[!htbp]
\fi
\centering
\includegraphics[width=0.88\textwidth]{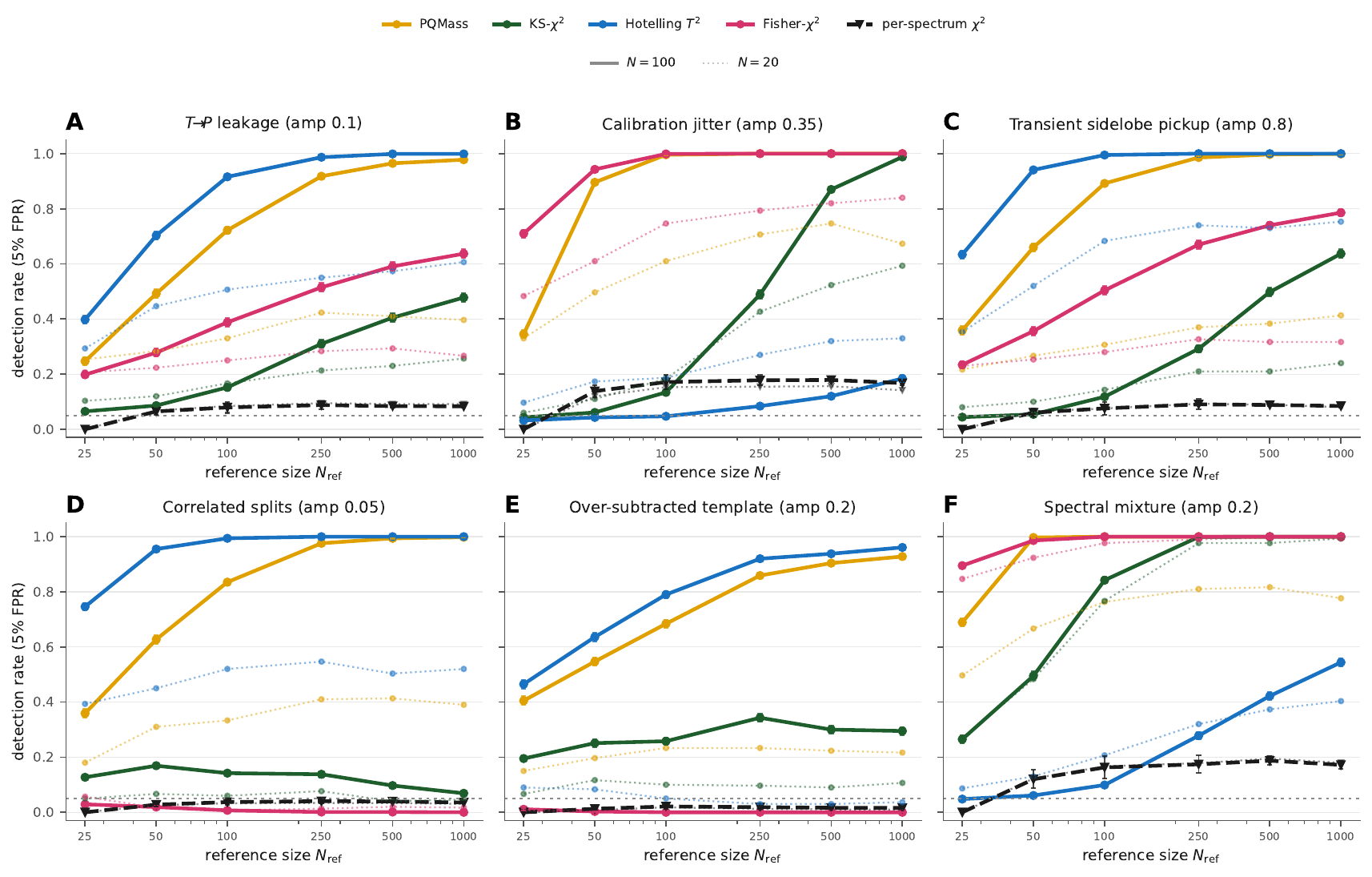}
\caption{\textbf{Detection rate versus reference ensemble size \(N_{\rm ref}\) for the SAT-motivated polarization null, at fixed representative contamination amplitudes and two data-ensemble sizes, \(N=100\) (bold) and \(N=20\) (dotted).} The \(N=100\) curves carry binomial error bars.
Panels show (\textbf{A}) temperature-to-polarization leakage, (\textbf{B}) calibration jitter, (\textbf{C}) transient sidelobe pickup, (\textbf{D}) correlated splits, (\textbf{E}) an over-subtracted template, and (\textbf{F}) a spectral mixture.}
\label{fig:fairref}
\end{figure}
}

\newcommand{\RealDataFigure}{%
\ifscienceformat
\begin{figure}[p]
\else
\begin{figure}[!tbp]
\fi
\centering
\includegraphics[width=\textwidth]{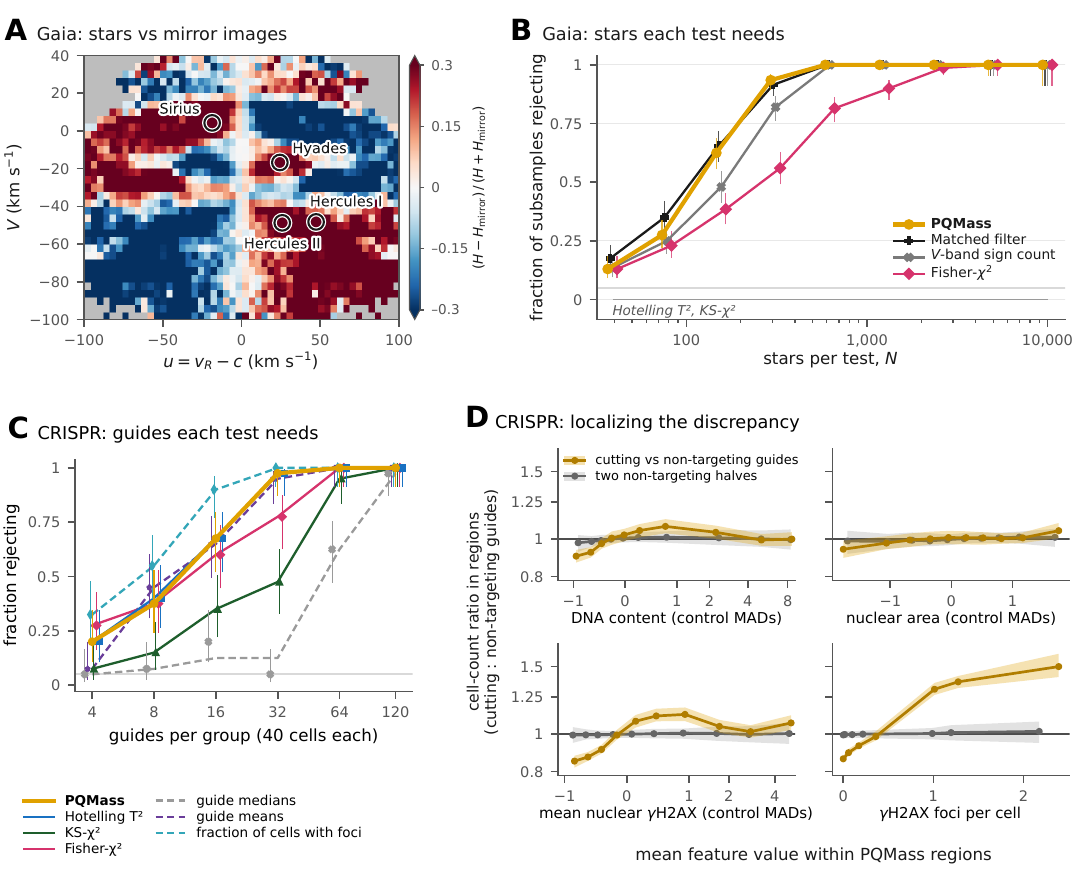}
\caption{\textbf{Detecting and localizing departures in Gaia and CRISPR.}
(\textbf{A}) Gaia DR3 velocity asymmetry within 200\,pc relative to radial
mirror images; circles mark published moving groups \cite{ramos2018}.
(\textbf{B}) \pqm{} approaches an
informed matched filter without using moving-group locations. Hotelling's
$T^2$ and KS-$\chisq$ miss the asymmetry; other tests need more stars for
high detection rates.
(\textbf{C}) Detection of differences between cells carrying cutting guides and non-targeting
controls as guides per group increase, with 40 cells per guide.
Solid lines denote \pqm{} and $\chisq$-type tests; dashed lines denote guide
summaries, including post hoc tests motivated by (D). \pqm{} outperforms guide medians and performs similarly to
Hotelling's $T^2$ and post hoc guide means.
(\textbf{D}) Localization: the same four-dimensional \pqm{} regions, grouped
by their mean feature values. Ratios above one indicate an excess of cells carrying cutting guides;
gray curves compare non-targeting halves. Curves and bands show median ratios
and 95\% intervals from guide resampling. The excess in regions with more
damage foci motivates the targeted, post hoc foci-fraction test (C, cyan) which is
the most sensitive. Such localization could guide targeted tests for
similar future experiments.
In B and C, curves show the fraction of subsamples rejecting at
$\mathrm{PTE}\le0.05$; error bars show 95\% intervals.}
\label{fig:realdata}
\end{figure}
}

\begin{document}
\maketitle
\begin{abstract}\bfseries
Unrecognized systematic errors can survive standard checks and undermine scientific discoveries.
Null tests and experimental controls can expose these failures, but averaging
measurements or reducing them to scalar statistics can hide important structure.
We propose ensemble null tests that compare multivariate distributions with
references from observations or simulations. We demonstrate their benefits 
across science. In cosmic microwave background simulations,
they detect systematic effects missed by
conventional $\chi^2$ tests.
In Gaia data, they detect velocity asymmetry as sensitively as an informed search. In a CRISPR screen, they expose a control mismatch in DNA damage and reveal which cellular profiles drive it.
As scientific datasets grow,
ensemble testing turns repeated measurements into a broader search for unknown
unknowns, strengthening the empirical foundation for scientific discovery.
\end{abstract}


\ifscienceformat\else
\begin{quote}
\small
\textit{``There are known knowns; there are things we know
we know. We also know there are known unknowns; that is to
say we know there are some things we do not know. But there
are also unknown unknowns---the ones we don't know we
don't know.''}
\par\hfill--- Donald Rumsfeld, U.S. Secretary of Defense, at a Pentagon briefing on February 12, 2002
\end{quote}
\fi

\noindent
As scientific datasets grow, statistical uncertainties shrink and
previously negligible systematic errors can become limiting. Recognized
effects with uncertain amplitudes or parameters constitute ``known unknowns'' that can often be modeled,
calibrated, or marginalized over. Unrecognized effects ---``unknown unknowns''---may instead lie
outside the assumed model and its uncertainty budget. Detecting such
failures requires tests that reveal inconsistencies without specifying
their physical cause in advance.

Scientists assess reliability using null tests and experimental controls.
Split-data null tests compare measurements expected to contain the same
signal, so their differences reveal inconsistencies. Controls provide
comparison measurements, such as untreated samples or common reference
materials. Repeated checks include control wells in microscopy
\cite{bray2016cellpainting}, MRI phantom scans \cite{stocker2005fmriqa}, metabolomics
\cite{sangster2006qc}, Raman
calibration standards \cite{lellinger2025raman}, and astronomical observations
divided by time, detector, or observing conditions
\cite{bicep2014,polarbear2014,choi2020,gatti2021,giblin2021}.
The shared statistical question is whether an ensemble of measurements
agrees with a suitable reference distribution. That reference can also
encode a physical symmetry---a rejected null then challenges that assumed symmetry.

Measurements can agree on average while differing in their variability or in how their features relate to one another.
In precision cosmology, for example, individual null measurements are typically
summarized by a $\chisq$ statistic and its probability to exceed (PTE)
\cite{bicep2014,polarbear2014,bicep2null,choi2020}.
This approach is simple and interpretable, but has three limitations.
First, inaccurate covariance estimates can misweight the data
and reduce sensitivity. Second, converting $\chisq$ to a PTE
requires a valid null calibration: analytic approximations can
be inaccurate for complex data, while Monte Carlo calibration
depends on the simulated null. Third, $\chisq$ retains only a
covariance-weighted magnitude: the distribution of these
scalar values discards directional and joint structure.
Additionally, coadding comparable observations into a few deep
nulls can erase observation-to-observation variation before
the statistic is formed. Together, these limitations can
miscalibrate rejection rates and hide detectable systematic
structure. The challenge is to retain sensitivity to diverse departures
while respecting each experiment's reference construction and calibration.

Examples from a few areas of science illustrate alternatives to a single $\chi^2$
summary, including diagnostics tailored to specific applications. MRI checks compare one-dimensional fluctuation distributions
with Gaussian noise \cite{stocker2005fmriqa}, while metabolomics checks inspect
control-sample clustering in two-dimensional principal-component projections
\cite{sangster2006qc}. In particle physics, machine-learning methods detect anomalies by comparing detector measurements with reference data collected under normal conditions
\cite{grosso2023dqm}.

We present a simple, broadly applicable approach to ensemble null testing that retains multivariate structure without prescribing the form of a possible departure or requiring neural network training. We introduce \pqm{} \cite{lemos2024pqm}, originally developed to assess generative models, to demonstrate what is gained by testing joint distributions rather than compressing each measurement vector to a single statistic or averaging across observations. We first isolate the effects of scalar compression in controlled examples. We then apply the framework to cosmic microwave background (CMB) $B$-mode simulations containing six injected systematics, radial-reflection symmetry in Gaia stellar velocities, and experimental controls in a CRISPR knockout screen. Together, these applications show how making fuller use of repeated measurements can expose departures hidden by conventional checks and reveal where their signatures lie within the data.

\subsection*{Ensemble null-test framework}

An ensemble null test asks whether repeated diagnostic measurements follow
an expected multivariate distribution. Each measurement is represented by a
vector of selected features, such as CMB bandpowers, stellar
velocities, or cellular profiles. The collection of measurement vectors forms a data ensemble
$\mathcal D$, which is compared with a reference ensemble $\mathcal R$
representing the null. In our applications, the reference comes from simulated null
measurements, held-out data, reflected data, or experimental controls.

\ifscienceformat\else
\SchematicFigure
\fi

Split-data diagnostics, commonly used in astronomical analyses, provide one
route to these vectors (Figure~\ref{fig:schematic}). For each repeated
measurement unit $i$, such as an observing night, two measurements are formed
from complementary subsets expected to contain the same signal:
\begin{equation}
m_i^A=s_i+n_i^A,
\qquad
m_i^B=s_i+n_i^B,
\end{equation}
where $n_i^A$ and $n_i^B$ include noise and split-dependent residuals.
Their difference,
\begin{equation}
d_i=m_i^A-m_i^B=n_i^A-n_i^B,
\end{equation}
cancels the shared signal. In CMB analyses, the splits can separate data by
observing time, detector group, or frequency band. Stellar velocities and
cellular profiles can instead be compared directly with a reference, without
subtracting a shared signal.

A conventional $\chisq$ statistic compresses a measurement vector $\mathbf{x}$ to
\begin{equation}
\chisq=(\mathbf{x}-\boldsymbol{\mu})^\top
\mathbf{C}^{-1}(\mathbf{x}-\boldsymbol{\mu}),
\end{equation}
where $\boldsymbol{\mu}$ and $\mathbf{C}$ are the reference mean and covariance.
This single number measures the covariance-weighted squared distance from
the mean but discards the direction of the departure. If $\mathbf{x}$ is
formed by averaging observations, differences in their spread or shape can
be lost before $\chisq$ is even computed.
To assess how these information losses affect sensitivity, we compare ensembles
using Hotelling's $T^2$, which targets mean differences;
Fisher-combined $\chisq$, which aggregates unusually large individual
deviations; Kolmogorov--Smirnov (KS)-$\chisq$, which compares distributions
of covariance-weighted distances; and \pqm{} \cite{lemos2024pqm}, which probes the full joint
distribution. 
\pqm{} partitions the joint measurement space into regions and compares the
proportions of data and reference points in each (see fig.~\ref{fig:pqmass-method} for a diagrammatic explanation). These comparisons also
help locate the combinations of features contributing to a discrepancy.
Mathematical definitions of these tests are given in section~\ref{app:tests}.
We use \pqm{} as a representative
multivariate test, with additional methods compared in
section~\ref{app:multivariate-comparison}. We also consider bespoke tests designed for the Gaia and CRISPR examples.

Calibration determines how unusual an observed difference is under
the null. Ensemble comparisons also offer a practical advantage: when labels are exchangeable under the null, significance can be calibrated by permutation without an analytic approximation to the test statistic’s null distribution. We use this approach for the controlled examples and CMB comparisons, and for CRISPR by permuting whole guides while keeping their cells together. Gaia instead uses paired sign randomizations because each star supplies its own reflected reference.

\subsection*{Controlled distributional examples}

We first isolate four changes in the location or shape of a cloud of points
that clarify what each test can detect: a mean shift, radial inflation,
radial compression, and angular concentration, illustrated in two dimensions
in Figure~\ref{fig:rung-illustration}A--D. For the detection tests, we draw
two clouds of 100 points from the same distribution. Each point represents
nine measurements from the CMB simulations described below. We apply the
same centering and rescaling to both clouds, using a fixed reference mean
and covariance to remove differences in scale and linear correlations.
We then shift or reshape one cloud while keeping the other as an unchanged
reference. Mathematical definitions are given in
section~\ref{app:controlled}.

\ifscienceformat\else
\ControlledExamplesFigure
\fi

\emph{Mean shift.}
We translate the cloud in a common direction. This isolates a location change
and therefore favors Hotelling. Pooled centering places the two means
symmetrically about their common center, making KS-$\chisq$ nearly insensitive
to the shift.

\emph{Radial inflation.}
We increase every point's distance from the center by the same factor while
preserving its direction. The larger distances make this change
readily detectable by Fisher and KS-$\chisq$.

\emph{Radial compression.}
We move each radius toward the sample mean radius, narrowing the radial spread
while preserving its mean. This also changes the covariance scale. It favors
the two-sided KS-$\chisq$, whereas Fisher's one-sided score moves away from
rejection.

\emph{Angular concentration.}
We move points toward the coordinate axes while approximately
preserving their distance from the center. This introduces higher-order
directional structure while largely preserving the lower-order features
targeted by the other tests.

Figure~\ref{fig:ladder}E--H shows Hotelling leading for the mean
shift, KS-$\chisq$ and Fisher for radial inflation, and KS-$\chisq$
for radial compression. \pqm{} strongly detects the angular-concentration
alternative that the mean- and distance-based tests largely miss, while
being less sensitive than KS-$\chisq$ for radial compression.
Targeted tests can therefore be more sensitive when their chosen feature
changes, whereas multivariate tests can also detect departures that
preserve that feature. The realistic applications below typically contain combinations of these different types of distributional shifts, and show how a general multivariate ensemble
test such as \pqm{} can achieve greater detection power when departures
involve structure missed by conventional summaries.

\subsection*{Safeguarding the search for primordial gravitational waves}

Searches for primordial CMB $B$-mode polarization require stringent
instrumental null tests. This faint signal could reveal an early-Universe
gravitational-wave background and probe inflationary physics
\cite{kamionkowski1997,zaldarriaga1997,seljak1997}, but must be separated
from Galactic foregrounds, gravitational lensing, noise, and systematic
contamination. Published BICEP/Keck and Planck analyses constrain the
tensor-to-scalar ratio to $r\lesssim0.03$ \cite{bicep2021,tristram2022}.

Repeated observations naturally supply null ensembles. BICEP/Keck use
multiple jackknife definitions \cite{bicep2null,bicep2021}; SPT-3G forms
null spectra from observation bundles \cite{spt3g2021}; and ACT uses
season, array, and detector splits \cite{act2020,choi2020,louis2025}.
The Simons Observatory small-aperture telescope (SAT) survey similarly
motivates comparable nulls per night, scan, or detector group \cite{so2019}.
Different split families must be tested separately or modeled jointly,
accounting for their distributions and shared data.

Our main comparisons use equal ensemble sizes
$N=|\mathcal D|=|\mathcal R|$, holding each spectrum's noise level fixed.
To test averaging at fixed total data, we separately partition each
ensemble's $N_{\rm obs}$ null maps into disjoint groups of $k$ units:
\begin{equation}
d_g^{(k)}=\frac{1}{k}\sum_{i\in g}d_i,
\label{eq:coadd-family}
\end{equation}
leaving
\begin{equation}
N_k=\left\lfloor\frac{N_{\rm obs}}{k}\right\rfloor
\label{eq:coadd-N}
\end{equation}
spectra per ensemble, with any remainder discarded. Larger $k$ produces
deeper spectra but fewer of them; $k=1$ retains observation-level nulls,
and $k=N_{\rm obs}$ gives a single fully coadded null.

\ifscienceformat\else
\SignaturesDetectionFigure
\fi

We simulate $B$-mode null maps on a
$20^\circ\times20^\circ$ flat-sky patch at
$2'$ resolution. Under the baseline null, we generate the two split noise maps
independently and subtract them to form $d_i$. We omit the shared sky signal
because it cancels exactly in the difference. Each split's noise
combines white noise, a residual $1/f$ component with $\ell_{\rm knee}=50$, a
$20\%$ log-normal modulation of the map depth, and scan-aligned striping.
The resulting noise is non-stationary across the patch and produces
correlated null bandpowers
(supplementary materials, section~\ref{app:noise},
fig.~\ref{fig:noisetypes}). Each CMB vector contains nine bandpowers
of the null map's auto-spectrum over $18\leq\ell\leq300$,
covering the recombination-bump scales relevant to primordial $B$-mode searches.

We inject one systematic at a time into the data ensemble while
keeping the reference uncontaminated. The upper plots of
Figure~\ref{fig:signatures} show six mechanisms spanning mean shifts,
excess scatter, reduced scatter, and changes in multivariate shape:
\begin{itemize}
\item \textbf{$T\!\to\!P$ leakage:} a split-dependent temperature
residual produces a mean shift concentrated at low $\ell$.
\item \textbf{Calibration jitter:} varying noise amplitudes produce
a scale mixture and altered scatter while preserving the mean input
noise power.
\item \textbf{Transient sidelobe pickup:} intermittent localized
contamination produces a mean shift and heavy tails.
\item \textbf{Correlated splits:} shared noise cancels in the
split difference, reducing the mean null power and scatter.
\item \textbf{Over-subtracted template:} excessive removal of
large-scale structure produces localized under-dispersion.
\item \textbf{Spectral mixture:} two populations with distinct noise
spectra, normalized to a common unweighted full-range spectrum sum,
produce an analyzed-band mean tilt and multivariate structure.
\end{itemize}
The mathematical definitions of these injections are given in
supplementary materials, section~\ref{app:noise}.
Contamination realizations are assumed independent across observing units.
For each scenario, we sweep the contamination amplitude from small values to values at which the most sensitive tests reject the null in nearly all mock experiments. At each
amplitude, we estimate detection rates using 500 independent experiments, each with newly generated reference and data ensembles.

The lower plots of Figure~\ref{fig:signatures} compare detection
rates at $N=100$ and $20$. Hotelling's $T^2$ remains near its nominal
rejection rate for calibration jitter, while Fisher and the per-spectrum
$\chisq$ respond weakly to correlated splits and underdetect
over-subtraction. \pqm{} has substantial power across all six
systematics. These results demonstrate the value of testing the full multivariate distribution---beyond
ensemble means and covariance-weighted distances.
Section~\ref{app:coverage} provides confirmation of false-positive control
and agreement between calibration methods.

Targeted tests can be more sample-efficient. At $N=20$, Fisher
outperforms the other tests shown for calibration jitter and spectral
mixture, which inflate the individual $\chisq$ values it aggregates;
at $N=100$, it matches \pqm{} on these scenarios over the amplitudes
studied. Nevertheless, multivariate methods provide a single, agnostic test sensitive
to a broad range of departures, without specifying their form in advance.

Other multivariate ensemble methods, including maximum mean discrepancy
(MMD) and energy-distance tests \cite{Gretton2012,Sejdinovic2013}, are
compared in the supplementary materials (section~\ref{app:multivariate-comparison},
fig.~\ref{fig:multivariate-comparison}).
MMD variants are competitive with \pqm{} and can be more sensitive
in some regimes, while their relative performance depends on the
systematic, contamination amplitude, and ensemble size.

\ifscienceformat\else
\CoadditionComparisonFigure
\fi

Coadding before testing can hide detectable contamination.
Figure~\ref{fig:twosplit} compares \pqm{} on an $N=100$ ensemble with a
fully coadded $\chisq$ using the same target observations. For the
observation-varying systematics studied here, ensemble testing reaches
50\% detection at contamination amplitudes as low as one-eighth those
required by the coadded test.

Supplementary tests examine the roles of measurement count, coadding depth,
and reference-ensemble size (section~\ref{app:analysis-choices}).

Coadding erases observation-to-observation structure, while $\chisq$
summarizes the remaining data by a covariance-weighted distance.
Coaddition can improve sensitivity to residuals that remain coherent across
observations. When contamination varies between observations, however,
ensemble testing can reveal what averaging conceals, using information
already present in the data.


\subsection*{Revealing broken symmetry in Gaia stellar motions}

A reference ensemble can encode a physical symmetry. In an axisymmetric
steady state, stars moving toward and away from the Galactic center should
have matching distributions of rotational velocity. Reversing radial motions
should therefore leave the joint velocity distribution unchanged. Known
moving groups---concentrations of stars with similar velocities---break this
symmetry \cite{dehnen1998,ramos2018}, providing a positive control.

\ifscienceformat\else
\RealDataFigure
\fi

We compare the radial and rotational velocities of Gaia DR3 stars within
200\,pc of the Sun \cite{gaiadr3} with a reference that reverses radial motion
about a center fixed on held-out stars, leaving rotation unchanged.
Figure~\ref{fig:realdata}A shows the excesses and deficits of stars relative
to this reflected reference. Star-wise sign flips calibrate the tests.
To limit sensitivity to uncertainty in this center, we restrict sign flips
to keep the mean radial motion nearly fixed. This targets asymmetry beyond
a bulk radial shift; Hotelling's $T^2$, which compares means, is therefore
insensitive by design. The reflection preserves the pooled
covariance-weighted distances, making our pooled KS-$\chisq$ insensitive.
Sample selection, test implementation, and calibration details are given in
section~\ref{app:gaia}.

Figure~\ref{fig:realdata}B shows \pqm{} closely tracking the detection curve
of an informed matched filter centered on published moving-group positions
\cite{ramos2018}, while detecting the asymmetry with fewer stars than the
other tests that lack prior group locations. The strongest of
these directly tests radial sign symmetry by comparing inward and outward counts within
rotational-velocity bins \cite{bowker1948}. Starting from the expected symmetry alone, \pqm{} approaches the sensitivity
of a search supplied with the locations of the moving groups. Detecting
asymmetry with fewer stars, without having to anticipate its form, could
enable finer spatial maps of where this symmetry breaks down, helping trace
the Galactic dynamics responsible.

\subsection*{Exposing DNA-damage signatures in CRISPR screens}

CRISPR knockout screens probe gene function by disrupting genes and measuring
how cells change. A guide RNA directs the DNA-cutting protein Cas9 to a
particular DNA sequence. Non-targeting guides control for the screening
procedure but do not direct cutting. A background response to cutting can
therefore be mistaken for an effect of disrupting the gene itself.

We reanalyze per-cell measurements from Funk et al.'s imaging-based screen of
cultured human (HeLa) cells \cite{funk2022}. We compare cells carrying non-targeting guides with cells carrying cutting guides
selected to target loci outside genes expressed in these cells.
We use DNA content and nuclear area, together with two measurements of
$\gamma$H2AX, a marker of the DNA-damage response: its mean intensity in the
nucleus and the number of localized spots, or foci. Guides were selected using genomic and gene-expression information, independently of the cellular measurements being tested. We calibrate the comparison by shuffling group labels
among whole guides, keeping the cells carrying each guide together.
Full implementation details are given in section~\ref{app:cutting}.

Figure~\ref{fig:realdata}C compares \pqm{} with $\chisq$-type tests (solid
lines) and additional summaries (dashed lines). Our guide-median benchmark, motivated by the original study \cite{funk2022},
compresses the cells carrying each guide to one median value per feature. \pqm{} reaches 50\% detection
with about 11 guides per group, compared with 54 for the median.
Hotelling's $T^2$ and Fisher-$\chisq$ have similar detection thresholds,
whereas KS-$\chisq$ needs more guides. Hotelling's performance shows that
a test of mean differences is already sensitive to this mismatch.

\pqm{} compares the multivariate distribution, so it can detect changes in
individual features and their correlations. By grouping cells with similar
measurement profiles into regions, \pqm{} also helps identify which parts
of the population contribute to the mismatch. Figure~\ref{fig:realdata}D shows
which regions contain an excess of cells carrying cutting guides, using their average
feature values to describe the affected cellular profiles. These cells are enriched in regions with more foci and intermediate $\gamma$H2AX
intensity, while control halves remain near parity. 

This localization points to a more targeted summary. Figure~\ref{fig:realdata}C
also shows post hoc comparisons using the same subsamples and permutations
(dashed lines): guide means perform similarly to \pqm{}, while the fraction
of cells with foci reaches 50\% detection with about seven guides per group.
Nearly every guide has a median foci count of zero, obscuring the higher average fraction of cells with foci for cutting guides.

The mismatch persists after matching well composition and analyzing
each imaging date separately (section~\ref{app:cutting}).
\pqm{} thus detects this control mismatch without requiring the right
summary statistic to be chosen in advance, using about one-fifth as many
guides as the median-based benchmark and achieving sensitivity comparable
to guide means. It also identifies the cellular profiles contributing to
the mismatch, suggesting targeted tests for similar future experiments.

\subsection*{Implications for scientific reliability}

We have shown how testing distributions of repeated
measurements can reveal departures hidden by conventional
summaries. In CMB simulations, ensemble comparisons detect
contamination that averaging measurements suppresses or
conventional $\chi^2$ diagnostics detect poorly. In Gaia data, a
general test of velocity symmetry approaches the sensitivity of
a search informed by moving-group locations. In the CRISPR screen,
comparing cellular distributions exposes a control mismatch that
guide medians capture poorly, with localization identifying the
cellular profiles contributing to it. These examples connect
instrumental consistency, physical symmetries, and experimental
controls through a common statistical framework.

We thus recommend retaining measurements from
individual observations
and testing their multivariate distributions, particularly when the form of a possible departure is unknown.
This preserves information that can be lost when observations
are averaged or reduced to a scalar statistic.
No single statistic is uniformly most sensitive, and targeted
tests remain valuable when a particular departure is anticipated.
Their complementary responses can help distinguish changes in
the mean, scatter, tails, or joint structure.
Depending on the
application, the discrepancy may reflect instrumental contamination,
an inadequate experimental control, or physical structure excluded
by the assumed symmetry. 

Beyond detection, \pqm{} can show which regions of the data contribute to a mismatch. In the CRISPR example, localization motivates a simple foci-based summary that is more sensitive than the general test. This progression from broad detection to a focused diagnostic can guide follow-up in independent data or targeted tests for future experiments.

The framework is relevant to a broad range of settings, including
CMB map and time-ordered-data pipelines, weak-lensing surveys
\cite{gatti2021,giblin2021},  gravitational-wave
time-slide backgrounds \cite{was2010,usman2016}, and Raman calibration  \cite{lellinger2025raman}.
The repeated observations that give modern experiments their
precision also contain information about whether their assumptions
hold. Ensemble null tests make fuller use of that information, broadening the search for unknown unknowns and strengthening the empirical foundation for scientific discovery.


\ifscienceformat
\SchematicFigure
\ControlledExamplesFigure
\SignaturesDetectionFigure
\CoadditionComparisonFigure
\RealDataFigure
\fi

\clearpage
\bibliography{refs}
\bibliographystyle{sciencemag}
\section*{Acknowledgments}
We thank Simone Aiola for insightful discussions and comments on the CMB example, and Laurence Perreault-Levasseur for insightful discussion on PQMass.
We also thank Zeeshan Ahmed, Zachary Atkins, Susanna Azzoni, Yuji Chinone, Adriaan
Duivenvoorden, Jo Dunkley, Mathew Madhavacheril, and Suzanne Staggs
for fruitful discussion on CMB applications. The computations in this work were performed on the Rusty
cluster at the Flatiron Institute, 
Simons Foundation.
This work has made use of data from the European Space Agency (ESA) mission
\textit{Gaia} (\url{https://www.cosmos.esa.int/gaia}), processed by the \textit{Gaia}
Data Processing and Analysis Consortium (DPAC,
\url{https://www.cosmos.esa.int/web/gaia/dpac/consortium}). This research has made use
of the VizieR catalogue access tool, CDS, Strasbourg, France.

\paragraph*{Software}
The analysis uses
\textsc{NumPy} 1.24 \cite{harris2020},
\textsc{SciPy} 1.10 \cite{virtanen2020},
\textsc{Matplotlib} 3.7 \cite{hunter2007},
\textsc{pixell} 0.28 \cite{pixell} (with the \textsc{ducc0} FFT backend),
PQMass 0.6.3 \cite{lemos2024pqm} (which uses PyTorch), and
scikit-learn 1.2 \cite{Pedregosa2011}, and, for the Gaia example,
\textsc{gaiadr3-zeropoint} 0.0.5 \cite{lindegren2021zp}.
The additional comparison also uses the MMDAgg reference implementation
(\textsc{mmdagg} 1.0.0; section~\ref{app:multivariate-comparison}).

\paragraph*{Use of generative AI tools.}
Portions of the analysis code and manuscript text were written with
the assistance of Anthropic's Claude Code (Opus 4.8--5.5 and Fable 5) and OpenAI's ChatGPT and
Codex (Sol 5.6 and Astra 6).
These tools were operated under the corresponding author's direction.
All methods, results, and text were reviewed and verified by the
authors, who take full responsibility for the content of this work.

\paragraph*{Funding:}
This work was supported by the Simons Foundation and by the National
Science Foundation under Cooperative Agreement PHY-2525568 (The NSF AI
Institute for Artificial Intelligence and Fundamental Interactions,
\url{http://iaifi.org/}).

\paragraph*{Author contributions:}
A.E.B. and D.N.S. conceived the study. A.E.B. developed and implemented
the methods, performed the analyses, prepared the
figures, and drafted the
manuscript. Both authors revised and approved it.

\paragraph*{Competing interests:}
The authors declare that they have no competing interests.

\paragraph*{Data and materials availability:}
All code used to generate the simulations, perform the analyses and
produce the figures, together with the simulated bandpower ensembles and
data underlying all results and figures in the main text and supplementary
materials, will be publicly available at Zenodo \cite{ensnull_zenodo} upon acceptance.
A private access link will be provided for peer review. Fig.~\ref{fig:realdata}A--B
uses public data: Gaia DR3 from the ESA Gaia archive \cite{gaiadr3,katz2023rv}
and cluster members \cite{hunt2024clusters} from CDS VizieR (J/A+A/686/A42). Fig.~\ref{fig:realdata}C--D uses the
single-cell features of the optical pooled screen of \cite{funk2022} (Harvard Dataverse, doi:10.7910/DVN/VYKTI5). All other data
needed to evaluate the conclusions in the paper are present in the paper or
the supplementary materials.

\subsection*{Supplementary materials}
General methods\\
CMB simulations and supplementary analyses\\
Gaia and CRISPR applications\\
Figs. \ref{fig:pqmass-method} to \ref{fig:fairref}\\
References \bibnumber{fisher1925}--\bibnumber{winkler2014}
\clearpage
\renewcommand{\thefigure}{S\arabic{figure}}
\renewcommand{\thetable}{S\arabic{table}}
\renewcommand{\theequation}{S\arabic{equation}}
\renewcommand{\thepage}{S\arabic{page}}
\renewcommand{\thesubsection}{S\arabic{subsection}}
\renewcommand{\thesubsubsection}{\thesubsection.\arabic{subsubsection}}
\setcounter{figure}{0}
\setcounter{table}{0}
\setcounter{equation}{0}
\setcounter{page}{1}
\setcounter{subsection}{0}
\setcounter{subsubsection}{0}
\begin{center}
\section*{Supplementary Materials for\\ \scititle}
Adrian E. Bayer$^\ast$ and David N. Spergel\\
\small$^\ast$Corresponding author. Email: abayer@mit.edu.
\end{center}
\label{supplement-start}
\subsubsection*{This PDF file includes:}
General methods\\
CMB simulations and supplementary analyses\\
Gaia and CRISPR applications\\
Figs. \ref{fig:pqmass-method} to \ref{fig:fairref}\\
References are included in the main article.
\newpage
\ifscienceformat\else
\renewcommand{\topfraction}{0.9}
\renewcommand{\bottomfraction}{0.85}
\renewcommand{\textfraction}{0.08}
\renewcommand{\floatpagefraction}{0.85}
\setcounter{topnumber}{2}
\setcounter{bottomnumber}{2}
\setcounter{totalnumber}{3}
\fi
\subsection*{General methods}
\subsection{Ensemble tests and calibration}
\label{app:tests}
\label{app:hyperparams}

The controlled and CMB comparisons use four global ensemble tests---
Fisher-combined $\chisq$, Kolmogorov--Smirnov (KS)-$\chisq$,
Hotelling's $T^2$, and \pqm{}. The CMB comparisons additionally show
the traditional per-spectrum $\chisq$ diagnostic.
These methods differ in the features they measure and their calibration.
Reference construction and calibration for Gaia and CRISPR are
described in sections~\ref{app:gaia} and~\ref{app:cutting}.
Additional multivariate tests are described and compared in
section~\ref{app:multivariate-comparison}. 
Unless otherwise stated, we evaluate the tests at
a nominal $5\%$ rejection level.

\emph{Per-spectrum $\chisq$.}
For each bandpower vector $\mathbf{x}_i$, we compute
\begin{equation}
q_i =
(\mathbf{x}_i-\widehat{\boldsymbol{\mu}})^\top
\widehat{\mathbf{C}}^{-1}
(\mathbf{x}_i-\widehat{\boldsymbol{\mu}}),
\end{equation}
the squared Mahalanobis distance from the null mean. A dedicated Monte
Carlo (MC) calibration suite is divided into two subsets of size 250: the
first estimates the mean $\widehat{\boldsymbol{\mu}}$ and covariance
$\widehat{\mathbf{C}}$, while the second supplies an out-of-sample
empirical null distribution of $q$. Each spectrum is assigned an
upper-tail PTE,
\begin{equation}
\mathrm{PTE}_i =
\frac{1+\sum_{s=1}^{N_q}\mathbb{I}(q_s\geq q_i)}
{N_q+1},
\end{equation}
where $\mathbb{I}$ is the indicator function and $N_q$ is the number
of held-out reference spectra used to tabulate the null distribution. Spectra with $\mathrm{PTE}<0.05$ are flagged, and we report the
mean flagged fraction within each experiment.
This is a per-spectrum
diagnostic, not a global test of the ensemble, and does not aggregate evidence across spectra. We use it as a point of reference for the ensemble methods.

\emph{Fisher-combined ensemble $\chisq$.}
To construct a global test from the per-spectrum $\chi^2$ values, we use Fisher's
statistic
\begin{equation}
\mathcal{F}=-2\sum_{i=1}^{N}\ln\mathrm{PTE}_i
\end{equation}
\cite{fisher1925}. 
As this is an ensemble test, we do not use any additional external calibration sets, but use $\mathcal{R}$ to fit the mean and covariance.

We also investigated estimating
$\widehat{\boldsymbol{\mu}}$ and $\widehat{\mathbf{C}}$ from the pooled
$\mathcal{D}\cup\mathcal{R}$, and from one half of $\mathcal{R}$ with the ranks
taken against the other; at matched simulation budget we found the $\mathcal{R}$ alone approach
to be the most powerful across all the systematics considered here. We consider scaling with the size of the reference ensemble in supplementary materials, section~\ref{app:analysis-choices}.
If the PTEs were independent and exactly uniform
under the null, $\mathcal{F}$ would follow a $\chisq_{2N}$ distribution---however, in general, this does not hold and we thus calibrate numerically.  Fisher's
method aggregates upper-tail evidence efficiently but is not designed to detect
ensembles that fluctuate systematically less than expected.

\emph{Ensemble KS-$\chisq$.}
This test compresses each bandpower vector to its covariance-weighted
squared distance within the pooled ensemble. Let
$\mathcal{P}=\mathcal{D}\cup\mathcal{R}$, with pooled mean
$\widehat{\boldsymbol{\mu}}_{\mathcal P}$ and covariance
$\widehat{\mathbf{C}}_{\mathcal P}$. We form
\begin{equation}
q_i =
(\mathbf{x}_i-\widehat{\boldsymbol{\mu}}_{\mathcal P})^\top
\widehat{\mathbf{C}}_{\mathcal P}^{-1}
(\mathbf{x}_i-\widehat{\boldsymbol{\mu}}_{\mathcal P}),
\end{equation}
and compare the empirical distributions of $q$ in the two ensembles
using the two-sample Kolmogorov--Smirnov statistic
\cite{kolmogorov1933,smirnov1948}
\begin{equation}
D_{\rm KS} =
\sup_q
\left|
\widehat F_{\mathcal D}(q)
-\widehat F_{\mathcal R}(q)
\right|,
\end{equation}
where $\widehat F_{\mathcal D}$ and $\widehat F_{\mathcal R}$ are the
empirical cumulative distribution functions of $q$ in the two
ensembles.
Because relabeling leaves the pooled sample unchanged, its mean and
covariance are invariant under permutation. Unlike the upper-tail
Fisher combination, KS-$\chisq$ responds to changes in either
direction, including excess power and under-dispersion. It retains the
full distribution of covariance-weighted spectral sizes but discards
directional information in bandpower space. Because pooled centering
places a pure mean shift on opposite sides of the common center (for
comparable ensemble sizes),
KS-$\chisq$ has little power against that alternative for an
approximately symmetric null. Conceptually, KS-$\chisq$ is related to
inspecting the distribution of $\chisq$ PTEs across a null suite,
here applied within a repeated split family and compared directly
with a reference ensemble.

\emph{Hotelling's $T^2$.}
Hotelling's statistic \cite{hotelling1931} targets a difference
between the ensemble means. Let
$n_{\mathcal D}=|\mathcal{D}|$ and
$n_{\mathcal R}=|\mathcal{R}|$, with means
$\overline{\mathbf{x}}_{\mathcal D}$ and
$\overline{\mathbf{x}}_{\mathcal R}$ and pooled within-group sample
covariance $\mathbf{S}_p$. The classical statistic is
\begin{equation}
T^2 =
\frac{n_{\mathcal D}n_{\mathcal R}}
     {n_{\mathcal D}+n_{\mathcal R}}
(\overline{\mathbf{x}}_{\mathcal D}
-\overline{\mathbf{x}}_{\mathcal R})^\top
\mathbf{S}_p^{-1}
(\overline{\mathbf{x}}_{\mathcal D}
-\overline{\mathbf{x}}_{\mathcal R}).
\end{equation}
For computational efficiency we replace $\mathbf{S}_p$ with the sample covariance of the pooled union $\mathcal{D}\cup\mathcal{R}$ (with a small diagonal loading, $10^{-10}\mathbb{I}$, added before inversion): the union is invariant under relabeling, so this matrix is inverted once per experiment rather than once per permutation. Without diagonal loading, and when the covariance matrices are invertible, the resulting quadratic form at fixed ensemble sizes is a strictly increasing function of the classical $T^2$ (by the Sherman--Morrison identity; \cite{sherman1950}), yielding identical permutation ranks and PTEs. The small loading regularizes our numerical implementation; permutation calibration applies directly to the statistic implemented.
Hotelling's test is powerful against
coherent mean shifts but does not directly test changes in ensemble
scatter or higher-order shape.

\textbf{\pqm{}.}
\pqm{} \cite{lemos2024pqm} is a nonparametric multivariate
two-sample test, recently applied in cosmological contexts \cite{bayer2025robust}.
It compares the probability mass that the two ensembles assign to regions
of measurement space, estimated by the fractions of their vectors in each
region (Fig.~\ref{fig:pqmass-method}). This retains joint structure that
mean- or distance-based compression can discard, without requiring a
Gaussian reference model or a full covariance matrix.

\ifscienceformat\else
\PQMassMethodFigure
\fi

For each tessellation, we standardize each coordinate using the pooled
data and reference mean and standard deviation, and randomly select
$K_{\rm PQM}$ seed vectors from the two ensembles. Each standardized
vector is assigned to its nearest seed in Euclidean distance, defining
non-overlapping Voronoi regions $V_j$. The seed vectors define the regions
and are then excluded from the counts. Writing $\mathcal D'$ and
$\mathcal R'$ for the remaining vectors, the counts are
\begin{equation}
d_j=\sum_{\mathbf x\in\mathcal D'}\mathbb I(\mathbf x\in V_j),
\qquad
r_j=\sum_{\mathbf x\in\mathcal R'}\mathbb I(\mathbf x\in V_j).
\label{eq:pqm-counts}
\end{equation}
Let $m_{\mathcal D}=|\mathcal D'|$ and
$m_{\mathcal R}=|\mathcal R'|$ denote the sample sizes after seed removal
for this tessellation. Under a common regional probability, the pooled
mass estimate and expected counts are
\begin{equation}
\widehat p_j=\frac{d_j+r_j}{m_{\mathcal D}+m_{\mathcal R}},
\qquad
E_{\mathcal D j}=m_{\mathcal D}\widehat p_j,
\qquad
E_{\mathcal R j}=m_{\mathcal R}\widehat p_j.
\label{eq:pqm-expected}
\end{equation}
The Pearson statistic for tessellation $t$ is
\begin{equation}
Q_t=\sum_{j:\,d_j+r_j>0}
\left[
\frac{(d_j-E_{\mathcal D j})^2}{E_{\mathcal D j}}
+\frac{(r_j-E_{\mathcal R j})^2}{E_{\mathcal R j}}
\right].
\label{eq:pqm-pearson}
\end{equation}
Regions empty in both ensembles are omitted. The expected counts account
for unequal sample sizes: the comparison concerns regional fractions,
not equality of raw counts. For equal remaining sample sizes, each
region contributes $(d_j-r_j)^2/(d_j+r_j)$.

For the main detection comparisons, the number of regions follows
\begin{align}
n_{\rm min}&=\min(|\mathcal D|,|\mathcal R|),\\
K_{\rm PQM}
&=\min\!\left[20,
\max\!\left(2,\left\lfloor n_{\rm min}/4\right\rfloor\right)\right].
\end{align}
We average over $L$ random tessellations,
\begin{equation}
T_{\rm PQM}=\frac{1}{L}\sum_{t=1}^{L}Q_t,
\label{eq:pqm-average}
\end{equation}
to reduce sensitivity to a particular partition. We use $L=10$. Larger values of $T_{\rm PQM}$ indicate
greater discrepancies in regional mass. We calibrate this averaged
statistic by the permissible relabelings described below and in the
application-specific sections, rather than assigning it a
$\chisq_{K_{\rm PQM}-1}$ distribution. The tessellations and counts are
recomputed at each relabeling. Pooled standardization is label-invariant,
and tessellation randomness is drawn independently of the relabeling,
preserving exchangeability when the relabeling scheme is valid under
the null.

The implementations of Hotelling's $T^2$, KS-$\chisq$, and the
Fisher combination used here require an estimate of the
$\Nbin$-dimensional covariance. When the dimension is too large
relative to the number of measurements used to estimate it, the
unregularized sample covariance is singular, requiring
regularization or dimension reduction. \pqm{} avoids covariance
inversion and remains computable when $N\ll\Nbin$
\cite{bayer2025robust}. Its statistical power still depends on
ensemble size, dimension, representation, and the alternative
being tested.

For the CMB comparisons, the per-spectrum $\chisq$ diagnostic uses
held-out null simulations to calibrate its PTE. The four ensemble tests compare
$\mathcal D$ directly with $\mathcal R$ using permutation calibration:
we pool the ensembles, repeatedly relabel their members while holding
the two sizes fixed, and rank the observed statistic among the
permuted values. Calibration of the additional multivariate tests,
including the separate MMDAgg procedure, is described in
section~\ref{app:multivariate-comparison}. 
Concretely, with $B=500$ relabelings, the
permutation PTE for a statistic $T$ is
\begin{equation}
{\rm PTE}_{\rm perm}=
\frac{1+\sum_{b=1}^{B}\mathbb I(T_b\geq T_{\rm obs})}{B+1}.
\end{equation}
The add-one correction prevents zero PTE \cite{phipson2010}. 
Permutation calibration is valid when the ensemble labels
are exchangeable under the null, as for independent
measurements drawn from a common distribution. The resulting
test controls the false-positive rate at or below the nominal
level at finite ensemble size, without a large-$N$
approximation or a separately simulated rejection threshold.
The reference $\mathcal R$ may consist of null simulations
or suitably comparable, disjoint observations. In the latter
case, calibration need not invoke an explicit noise model,
but exchangeability remains an assumption about the data.
Equal marginal distributions or disjoint data subsets alone
do not guarantee exchangeability when measurements are
dependent.

For other applications, the unit of replication and the permissible
permutations must follow the experimental design. Cells within
a well, repeated measurements of one specimen, or observations
within a shared experimental run cannot automatically be treated
as independent exchangeable replicates. Valid relabelings must
respect relevant dependence, matching, and blocking; otherwise,
a different calibration strategy is required. Preprocessing that
depends on the ensemble labels must be included in the permutation
procedure. Normalizing data and reference separately can also
remove discrepancies that the test is intended to detect.

Calibration under the exact null, and agreement between permutation
and external-suite Monte Carlo calibration for the four main-text
ensemble tests, are examined in section~\ref{app:coverage}.
Null rejection rates for the additional comparison are reported
in section~\ref{app:multivariate-comparison}.

Computational cost depends on the statistic and its implementation.
The pooled covariance used by Hotelling's $T^2$ and KS-$\chisq$
is invariant under relabeling and is inverted once per experiment;
KS-$\chisq$ also reuses the squared distances. The Fisher combination
re-estimates the reference mean and covariance, and the \pqm{}
implementation redraws its tessellations, at each relabeling.
The additional benchmark compares computational costs in
section~\ref{app:multivariate-comparison}.

\subsection*{CMB simulations and supplementary analyses}

\subsection{Simulation and injection details}
\label{app:noise}

\subsubsection{SAT-motivated baseline null}
\label{app:baseline}

We use a controlled SAT-motivated statistical model rather than an
end-to-end polarization simulation. Each split map is generated as a
scalar random field, and the power spectrum of their difference is
interpreted as the $B$-mode null spectrum. We do not simulate Stokes $Q/U$
maps or perform $E/B$ separation. The patch is unmasked and periodic, so the
bandpowers require neither a mode-coupling correction nor $E/B$
purification.

Our baseline assumes exact cancellation of the shared sky signal. Expected
residuals that survive a split, such as foreground emission in a frequency
split, must be included in the reference even for a perfect instrument.

Each split is generated independently on a $600\times600$ pixel grid
at $2'$ resolution, corresponding to a
$20^\circ\times20^\circ$ patch. Gaussian Fourier modes are
synthesized through $\ell=6000$; the map is the real part of the
inverse transform of complex white noise filtered by $\sqrt{C_\ell}$. Each split contains the following components.

\emph{White noise.}
An isotropic Gaussian field is drawn from the flat spectrum
$N_\ell=w$ ($w=0.002$ in the generator's internal units).

\emph{Residual $1/f$ noise.}
An independent Gaussian field is drawn from
\begin{equation}
N_\ell^{\rm red}\propto
1+\left(\ell/\ell_{\rm knee}\right)^{-\alpha},
\qquad
\ell_{\rm knee}=50,
\quad
\alpha=2.5,
\end{equation}
with $\ell$ floored at unity. The realized red-noise map is rescaled
to have half the rms of the white map before the two are added.
The choices $\ell_{\rm knee}=50$ and $\alpha=2.5$ are
consistent with published pessimistic SAT polarization forecasts, which
adopt $\ell_{\rm knee}\simeq 50$ and
$-3.0\lesssim\alpha_{\rm knee}\lesssim-2.4$ under the opposite sign
convention \cite{so2019}.

\emph{Non-stationary depth.}
The white-plus-red map is multiplied by the mean-one log-normal field
\begin{equation}
A(\hat n)=\exp\!\left(\sigma_A\widehat g(\hat n)
-\frac{\sigma_A^2}{2}\right),
\qquad \sigma_A=0.2,
\end{equation}
where $\widehat g$ is a unit-variance Gaussian field obtained by
smoothing pixel white noise with a $150$-pixel ($5^\circ$) Gaussian
kernel. This gives approximately 20\% rms modulation in map amplitude
and is redrawn for every split.

\emph{Scan-aligned striping.}
A one-dimensional Gaussian profile is obtained by smoothing per-row
white noise with a $75$-pixel ($2.5^\circ$) Gaussian kernel. It is
broadcast along the orthogonal axis and added with rms equal to 10\%
of the white-plus-$1/f$ map rms. The stripe orientation is fixed, but
the profile is redrawn for every split.

\emph{Variance matching.}
Each completed map is rescaled by one map-wide factor to match the rms
of its white-base realization. The SAT-motivated and white-noise nulls
therefore have the same overall map variance but different angular
correlations and non-stationarity. Where a power normalization is required, we express it relative to
the flat white-noise level \(N_\ell^w=w\); the remaining injection
parameters are dimensionless.
Figure~\ref{fig:noisetypes} shows an example null map and the
corresponding bandpower ensemble.

\ifscienceformat\else
\BaselineNoiseFigure
\fi

\subsubsection{Controlled distributional transformations}
\label{app:controlled}

The controlled examples in main-text Figure~\ref{fig:rung-illustration} use
the nine-dimensional reference distribution from the SAT-motivated baseline
null, with $N=100$ vectors in each ensemble. Let $\boldsymbol\mu_0$ and
$\mathbf C_0$ denote the mean and covariance of this reference distribution,
estimated from $4\times10^4$ simulated null spectra, and define
\begin{equation}
\mathbf z=\mathbf L^{\top}(\mathbf x-\boldsymbol\mu_0),
\qquad
r=\lVert\mathbf z\rVert,
\qquad
\widehat{\mathbf z}=\mathbf z/r,
\end{equation}
where $\mathbf L$ is the lower-triangular Cholesky factor of
$\mathbf C_0^{-1}$ ($\mathbf L\mathbf L^{\top}=\mathbf C_0^{-1}$), so that
$\mathbf z$ has zero mean and unit covariance. The reference ensemble is
unchanged, while one transformation is applied to the data ensemble with
amplitude $a$. These fixed coordinates define the alternatives only; the tests
and their permutation calibration are described in section~\ref{app:tests}.

\emph{Mean shift.}
We translate the cloud along the all-ones direction,
\begin{equation}
\mathbf z\longrightarrow
\mathbf z+a\frac{\mathbf 1}{\sqrt{\Nbin}}.
\end{equation}

\emph{Radial inflation.}
We rescale $\mathbf z\longrightarrow(1+a)\mathbf z$, increasing every
radius while preserving direction.

\emph{Radial compression.}
We contract radii toward the sample mean radius $\bar r$,
\begin{equation}
\mathbf z\longrightarrow
\left[\bar r+(1-a)(r-\bar r)\right]\widehat{\mathbf z},
\qquad 0\leq a\leq1.
\end{equation}
This preserves the mean radius while reducing its spread and therefore
changes the covariance scale rather than providing a moment-matched
alternative.

\emph{Angular concentration.}
We preserve radius while moving each direction toward its nearest
signed coordinate axis. Defining
\begin{equation}
k(\mathbf z)=\arg\max_j|z_j|,
\qquad
\mathbf u(\mathbf z)=
\operatorname{sgn}(z_{k(\mathbf z)})\mathbf e_{k(\mathbf z)},
\end{equation}
we set
\begin{equation}
\widehat{\mathbf z}\longrightarrow
\frac{(1-a)\widehat{\mathbf z}+a\mathbf u(\mathbf z)}
{\lVert(1-a)\widehat{\mathbf z}+a\mathbf u(\mathbf z)\rVert}.
\end{equation}
This leaves every radius unchanged up to a small finite-sample recentering
in the implementation (percent-level or smaller changes in the saved runs).

\subsubsection{Injected systematics}
\label{app:injections}

The two split halves receive independent baseline-noise realizations
unless the injected scenario explicitly correlates them. A
contamination is applied to split $A$ relative to split $B$, or
directly to $d_i$, and is redrawn for every observing unit in
$\mathcal D$; $\mathcal R$ remains uncontaminated. The suite therefore
targets observation-varying failures that alter the distribution of
null spectra.

\emph{Calibration jitter.}
For every observing unit, split $A$ receives the log-normal
power rescaling
\begin{align}
g_c&\sim\mathcal N(0,\sigma_c^2),\\
\alpha_c&=\exp\!\left(g_c-\frac{\sigma_c^2}{2}\right),
\qquad
N_\ell^A\longrightarrow\alpha_c N_\ell^A,
\end{align}
so that $\langle\alpha_c\rangle=1$. This creates a scale mixture and
changes the ensemble scatter while approximately preserving the mean
noise power. The injection modulates only the noise power; a gain error
acting on the shared sky signal would additionally leave a coherent
residual, which we do not model.

\emph{Spectral mixture.}
Split $A$ draws with equal probability from two tilted spectra,
\begin{equation}
N^A_\ell\propto w\left(\frac{\ell}{100}\right)^{+a}
\quad\text{or}\quad
N^A_\ell\propto w\left(\frac{\ell}{100}\right)^{-a}.
\end{equation}

Each branch is normalized to the same unweighted $C_\ell$
sum over the full simulated multipole range. This normalization
does not enforce equal map variance, which weights the spectrum
by mode count, or equal power within the analysis band.
The injection therefore also changes the in-band mean power,
producing two populations separated along a scale-dependent
direction in bandpower space.

\emph{Temperature-to-polarization leakage.}
Leakage common to both halves cancels in $d_i$, so we model only a
split-dependent residual. A random field is added to split $A$ with
\begin{equation}
C_\ell^{\rm leak}=\epsilon^2
\left(\frac{\ell}{100}\right)^{2a}C_\ell^T,
\qquad
C_\ell^T=50\,w\left(\frac{\ell}{100}\right)^{-1},
\end{equation}
where $\epsilon$ is the split mismatch at $\ell=100$ and the fiducial
scale-dependence exponent is $a=-1/2$.

\emph{Transient sidelobe pickup.}
A Poisson number of localized events,
$N_{\rm blob}\sim\operatorname{Poisson}(1.5)$, is added to split $A$.
Each event has the form
\begin{equation}
\pm A_{\rm side}\,\sigma_n
\exp\!\left(-\frac{r^2}{2\sigma_{\rm pix}^2}\right),
\qquad
\sigma_{\rm pix}=8\ \text{pixels}\ (16'),
\end{equation}
where $\sigma_n$ is the baseline rms of split $A$ and the sign is
drawn independently. Centers are uniform over the grid, and events
crossing a boundary are truncated rather than wrapped. The Poisson
count makes the contamination intermittent and produces a
heavy-tailed ensemble.

\emph{Correlated splits.}
The two halves share a fraction $\rho$ of their noise,
\begin{align}
n_A &= \sqrt{\rho}\,n_c+\sqrt{1-\rho}\,n_A',\\
n_B &= \sqrt{\rho}\,n_c+\sqrt{1-\rho}\,n_B',
\end{align}
where the three fields on the right are independent baseline draws
with equal marginal power. The common component cancels in $d_i$,
reducing its variance by a factor $1-\rho$ relative to the
independent-half reference.

\emph{Over-subtracted template.}
A fraction $a$ of a smoothed copy of the null difference is removed,
\begin{equation}
d_i\longrightarrow d_i-a\,\mathcal G_\sigma[d_i],
\qquad \sigma\simeq L/8,
\end{equation}
where $\mathcal G_\sigma$ is a Gaussian low-pass operator and $L$ is
the map side length. This mimics an over-aggressive template or filter
that removes genuine large-scale fluctuations and produces localized
under-dispersion.

\paragraph*{\pqm{} configuration and sensitivity.}
The shared sample-size rule in section~\ref{app:hyperparams} gives
$K_{\rm PQM}=20$ for $N=100$ and $K_{\rm PQM}=5$ for $N=20$ in the
equal-size CMB comparisons. Both the external calibration bank and the
permutation procedure average the statistic over ten random tessellations,
so they differ only in their rejection threshold. The standalone
hyperparameter evaluations, which rank a statistic against an independently
drawn reference rather than against relabelings of its own sample, use
fifty tessellations. A degenerate \pqm{} evaluation is conservatively
assigned ${\rm PTE}=1$ and retained.

This configuration---the pooled standardization, the cell-count rule, and the
tessellation depth---was selected once, informed by a sweep over the injection
suite and by the package's recommendations, and held fixed across the CMB
injection scenarios and ensemble sizes; no per-scenario tuning is performed,
and the production power curves use fresh realizations generated after that
choice. To assess sensitivity to this choice, we varied $K_{\rm PQM}$ over $\{5,10,20,50\}$ at
eleven (scenario, amplitude) points at which no test is saturated. \pqm{}'s detection rate typically varies by $\sim0.1$ across that range
and is usually largest near the adopted $K_{\rm PQM}=20$, though in some
cases a different choice would have led to slightly better performance.

\subsection{Coverage and calibration validation}
\label{app:coverage}

We test false-positive control on 1000 uncontaminated experiments at both $N=100$
and $N=20$, with $\mathcal D$ and $\mathcal R$ drawn independently from the same
null generator, and then ask whether the calibration route affects the detection
rates reported in the main text.

Figure~\ref{fig:coverage-pp} compares the achieved rejection probability
$\widehat P(p\leq\alpha)$ with the nominal rate $\alpha$. All four tests sit
within about one percentage point of nominal at $N=100$, against a $95\%$
binomial uncertainty of $1.4$ points; at $N=20$ KS-$\chisq$ becomes conservative,
reaching $2.8\%$ at $\alpha=0.05$. We find no material evidence of excess
rejection over $0.01\leq\alpha\leq0.5$, the bands being pointwise so that small
excursions across a scan of many $\alpha$ are expected. The
per-spectrum $\chisq$ flag fraction is $\simeq4\%$ at both sizes, conditional on
the single calibration bank whose held-out distances set the
$12/251\simeq4.8\%$ threshold. An external bank of independent clean ensemble
pairs, used to calibrate in place of permutation, agrees with the production
curves within binomial uncertainty and so provides an independent check of the
implementation. Figure~\ref{fig:coverage-rank} shows the corresponding PTE
distributions, including the expected discreteness at small $N$.

\ifscienceformat\else
\CoverageFigure
\PermutationRanksFigure
\fi

\ifscienceformat\else
\CalibrationComparisonFigure
\fi

False-positive control under the null does not by itself make the reported detection rates
route-independent: under a contaminated alternative $\mathcal D$ and
$\mathcal R$ are no longer exchangeable, and two tests of identical size may
differ in power. Figure~\ref{fig:calibgallery} therefore evaluates every
statistic twice on the same experiments, once against its permutation threshold
and once by ranking it against a bank of 1000 clean null universes under the
same add-one convention. The two routes agree to $0.009$ rms over the different combinations. Relabeling a contaminated pool does inflate the
permuted statistic's spread, and so raises the permutation threshold slightly,
but the effect remains small. For the four main-text tests, this comparison indicates that the
reported sensitivity differences are not driven by the choice
between these two calibration routes.

\subsection{Comparison of multivariate ensemble tests}
\label{app:multivariate-comparison}

We assess whether the sensitivity demonstrated with \pqm{} extends
to other multivariate tests by comparing it with energy distance
\cite{Sejdinovic2013}, a multiscale maximum mean discrepancy
(MMD) statistic \cite{Gretton2012}, MMDAgg \cite{Schrab2023},
and a classifier two-sample test (C2ST) based on trees
\cite{LopezPaz2017}.
We use the six CMB-motivated injections of
section~\ref{app:noise} with $N=100$.
Within this comparison, all methods receive the same simulated
data and reference vectors. Coordinates are standardized using
the pooled mean and sample standard deviation, computed without
using ensemble labels. Detection rates are estimated from 500
independent experiments per contamination amplitude at a nominal
rejection level of 5\%.
\pqm{} uses the implementation in section~\ref{app:hyperparams}.

\paragraph*{Additional statistics and calibration.}
Energy distance compares between-ensemble and within-ensemble
pairwise distances. For max-MMD, we use Gaussian kernels
$k_\sigma(\mathbf x,\mathbf y)=\exp(-\lVert\mathbf x-\mathbf y\rVert^2/2\sigma^2)$
at $\sigma\in\{0.25,0.5,1,2,4\}\,\sigma_{\rm med}$, where $\sigma_{\rm med}$
is the median off-diagonal pairwise distance in the pooled sample.
For each bandwidth, we compute the unbiased $\mathrm{MMD}^2$ estimator
\cite{Gretton2012} for the observed labeling and for each relabeling,
standardize it by its mean and standard deviation over these $B+1$
labelings, and take the maximum across bandwidths as the statistic.
Because this standardization is computed from the full set of labelings,
it treats the observed and permuted assignments symmetrically.
The pooled coordinate standardization and bandwidth scale are
invariant under relabeling. Energy distance and max-MMD use
$B=500$ permutations and the add-one PTE convention described
in section~\ref{app:tests}.

MMDAgg combines tests at ten bandwidths for each of the
Gaussian and Laplace kernels, with uniform weights and jointly
calibrated rejection thresholds \cite{Schrab2023}. We use version 1.0.0
of the authors' implementation with $B_1=B_2=2000$ resampling draws and
$B_3=50$ bisection steps, using permutations rather than the wild
bootstrap for equal ensemble sizes. The joint correction accounts for
testing multiple kernels and bandwidths; it is part of the method.
This procedure differs from ranking a single statistic using the
shared $B=500$ permutations.

C2ST uses the histogram-based gradient-boosted tree classifier of
scikit-learn (version 1.2.2) \cite{Pedregosa2011} with default settings
(100 boosting iterations, learning rate 0.1, at most 31 leaves per tree,
minimum leaf size 20). Its statistic is the mean held-out accuracy over stratified five-fold cross-validation. For the observed labeling and each of the $B=500$ permuted labelings, the stratified folds are rebuilt from that labeling, preserving equal numbers of vectors from each ensemble in every fold; only the random seed of the fold split is fixed within an experiment. The classifier is refit on every fold of every labeling, and the same add-one PTE convention is used.

Independent checks using 1000 uncontaminated experiments at $N=100$
give rejection rates of 6.1\% for \pqm{}, 4.7\% for max-MMD,
4.9\% for MMDAgg, 4.6\% for energy distance, and 5.7\% for C2ST.
Each lies within the pointwise 95\% binomial interval (3.6--6.4\%)
expected for nominal 5\% rejection.

\ifscienceformat\else
\MultivariateComparisonFigure
\fi

\paragraph*{Sensitivity.}
Figure~\ref{fig:multivariate-comparison} shows that all five
multivariate tests detect each of the six systematics,
approaching unit rejection rates at the largest amplitudes
tested. \pqm{} and the two MMD implementations generally
produce similar detection curves. Energy distance is less
sensitive to calibration jitter, while C2ST requires stronger
contamination in several scenarios. These differences do not
alter the central conclusion: broad sensitivity to systematic
contamination can be achieved with several multivariate
statistics. The results therefore support using \pqm{} as
a representative multivariate test in the main figures.

\paragraph*{Computational cost.}
At $N=100$, median runtimes over five timed runs
on one core are approximately $0.0042$ s for energy distance,
$0.013$ s for max-MMD, $2.8$ s for \pqm{}, and $2.9$ s for
MMDAgg, including each method's calibration.
The energy-distance and max-MMD implementations reuse pooled
distance or kernel matrices across relabelings, whereas this
\pqm{} implementation redraws its tessellations.
These timings characterize the implementations and calibration
budgets used here, including MMDAgg's different resampling
budget; they are not universal complexity rankings.
C2ST, which refits the classifier on every fold of every relabeling,
requires approximately $96$ s.

\subsection{Scalings}
\label{app:analysis-choices}

The sensitivity of an ensemble null test is governed by three quantities, each of
which we vary in turn. Section~\ref{app:coadding} varies the coadd depth $k$ at
fixed total data, so that deeper spectra are obtained at the expense of
their number, $N_k=\lfloor\Nobs/k\rfloor$. This corresponds to repartitioning a fixed
volume of data, and is the trade implicit in any conventional coadded
analysis; each depth is evaluated on independent realizations of the same
null model and data volume. Since that sweep varies depth and number simultaneously,
Section~\ref{app:ensemble-size} varies the number of spectra $N$ with the
construction and noise level of each held fixed; this corresponds to additional
comparable observing units rather than a repartitioning, and thereby isolates the
benefit of sampling the ensemble distribution more densely. Finally,
Section~\ref{app:refsize} varies the size of the reference ensemble $\Nref=|\mathcal{R}|$
alone at fixed data ensemble size $|\mathcal{D}|$, thereby distinguishing the simulation budget from
the data itself in cases where $\mathcal{R}$ is produced with simulations.

\subsubsection{Coadd depth}
\label{app:coadding}

The coadd construction in the main text connects the
observation-level ensemble continuously to a conventional fully coadded null, and
main-text Figure~\ref{fig:twosplit} compares the two endpoints on the same observations;
here we isolate the effect of the depth $k$ itself. Increasing $k$ yields deeper
spectra but fewer of them, and averages away mixtures, intermittency and other
observation-to-observation structure before any statistic is formed. For the depth sweep, the external
calibration is rebuilt at each depth from 500 null spectra constructed at that
same depth, so the sweep does not degrade the per-spectrum $\chisq$ baseline as it
proceeds.

The fully coadded $\chisq$ in main-text Figure~\ref{fig:twosplit} is
calibrated with an external suite of 400 independent null universes,
coadded identically to the target data: 200 estimate the mean and covariance
and 200 tabulate the null distribution.

\ifscienceformat\else
\CoaddDepthFigure
\fi

Figure~\ref{fig:apjl-coadding-tradeoff} shows that every test with initial power
loses sensitivity as the depth increases. Calibration jitter illustrates how
rapidly: \pqm{} falls from $99\%$ at $k=1$ to $43\%$ at $k=2$, and is consistent
with its false-positive rate by $k=5$. The structural blind spots persist
throughout---Hotelling remains near its null rate for calibration jitter, and
Fisher at or below its null rate for correlated splits and over-subtraction. This
sweep is not a prescription for an optimal depth, which depends on the noise and
systematic coherence of a given experiment; it quantifies the information lost in
the setup studied here.

\subsubsection{Ensemble size}
\label{app:ensemble-size}

Because the coadd sweep varies spectrum depth and ensemble size together, we also
vary $N$ with the construction and noise level of each spectrum held fixed. This
represents additional comparable observing units rather than a repartitioning of
fixed data, and so isolates the benefit of sampling the ensemble distribution
more densely.

\ifscienceformat\else
\EnsembleSizeFigure
\fi

Figure~\ref{fig:msweep} shows detection by the responsive tests improving steeply
with $N$ in all six scenarios at the chosen amplitudes; \pqm{}, for example,
typically rises from $15$--$30\%$ at $N=20$ to near saturation by $N=100$--$250$.
The per-spectrum $\chisq$ flag probability does not accumulate in the same way,
and increasing $N$ only narrows the realized uncertainty on the reported flag
fraction.

\subsubsection{Reference ensemble size}
\label{app:refsize}

This sweep holds the data-ensemble size fixed, at $N=100$ and $N=20$, and varies
the size $\Nref$ of the clean ensemble each test is compared against, isolating
the sensitivity each test draws from its reference. It begins at $\Nref=25$,
above the admissibility bound of both constructions: the held-out per-spectrum
reference splits $\Nref$ into disjoint fitting and tabulating halves and requires
$\Nref\geq2(\Nbin+2)=22$, whereas Fisher uses its reference whole and requires
only $\Nref\geq\Nbin+2=11$.

\ifscienceformat\else
\ReferenceSizeFigure
\fi

Figure~\ref{fig:fairref} shows all four tests gaining with $\Nref$, though they
saturate differently. \pqm{} and
Hotelling's $T^2$ rise steeply up to $\Nref\simeq N$ and largely saturate there,
whereas KS-$\chisq$ and Fisher continue to gain across the full axis: for
calibration jitter KS-$\chisq$ reaches $13\%$ at $\Nref=100$ against $99\%$ at
$1000$, and for $T\!\to\!P$ leakage Fisher rises from $39\%$ to $64\%$ over the
same interval. Fisher's continued gain has two sources, since
$\boldsymbol{\mu}$ and $\mathbf{C}$ are estimated from $\mathcal R$ alone: the
whitening metric and the per-spectrum PTE resolution both improve with $\Nref$,
the latter through the add-one floor $1/(\Nref+1)$ that caps how much any single
spectrum can contribute to $\mathcal F$. Growing the two roles separately
attributes most of the rise to the floor---$8$--$26\%$ of the gain from the
metric sample alone, against $54$--$72\%$ from the rank sample.

\subsection*{Gaia and CRISPR applications}


\subsection{Revealing broken symmetry in Gaia stellar motions}
\label{app:gaia}

\paragraph*{Sample and reference.}
We use Gaia DR3 stars with radial velocities \cite{gaiadr3} within 200\,pc
(zero-point-corrected parallax above 5\,mas \cite{lindegren2021zp}), parallax
over error above 10, RUWE below 1.4, radial-velocity error below
5\,km\,s$^{-1}$, template temperature below 8{,}500\,K and no duplicated-source
flag. Radial velocities are corrected for their magnitude trend
\cite{katz2023rv}. We exclude members of open clusters and compact co-moving
groups \cite{hunt2024clusters} to reduce dependence among co-natal stars,
while retaining the broad velocity-space structures used as the positive
control \cite{bovy2010}. A random 20\% of the remaining stars fixes the center
$c=-3.16\pm0.12$\,km\,s$^{-1}$, their mean Galactocentric radial velocity,
and is excluded from testing. The tested coordinates are $(u,V)$, where
$u=v_R-c$ and $V=v_\phi-245.6$\,km\,s$^{-1}$; the reference is $(-u,V)$.
Reflection preserves each star's position, rotational velocity and radial
speed relative to the fitted center.

\paragraph*{Tests and calibration.}
\pqm{} uses pooled standardization, ten tessellations and the sample-size
rule in section~\ref{app:hyperparams}; the three other main-text tests are
defined there as well. Additional comparisons use energy distance and eight
velocity-symmetry statistics, including a Bowker-type binned symmetry statistic
\cite{bowker1948} and a matched filter counting stars in 8\,km\,s$^{-1}$ disks
at four published moving-group positions listed by Ramos et al.
\cite{ramos2018}. Calibration swaps stars with their mirror images. To limit
sensitivity to uncertainty in $c$, sign flips $s_i=\pm1$ are drawn uniformly
among those satisfying
$|\sum_i s_i u_i-\sum_i u_i|\le0.05\lVert\mathbf u\rVert$.
This limits changes in mean radial motion to 5\% of the unconditioned
sign-flip standard deviation, targeting asymmetry beyond a bulk shift.
PTEs use 199 relabelings (999 at $N=10{,}000$). Calibration is approximate;
synthetic null checks using the real design give 4--6\% rejection at a
nominal 5\%.

\paragraph*{Detection power.}
We successively halve the sample size from 10{,}000 stars to 39, testing up
to 200 disjoint random subsamples per size at $\mathrm{PTE}\le0.05$.
Figure~\ref{fig:realdata}A shows $(H-H_{\rm mirror})/(H+H_{\rm mirror})$
in 4\,km\,s$^{-1}$ cells, gray below 20 stars and mirror images combined.
Figure~\ref{fig:realdata}B shows 95\% Wilson intervals for the rejection fractions.

\paragraph*{Controls and robustness.}
Random halves of unreflected stars do not reject in 10{,}000-star blocks
after multiple-testing correction. Sign-symmetric controls (random signs
assigned to $|u|$) also do not reject at this size. Detection persists when signs are flipped for
whole 50-pc spatial cells and when tessellation seeds are changed.
\pqm{} retains its sensitivity advantage over the uninformed comparisons. The informed matched filter performs better under some alternative calibration choices. Calibration assumes independent stars; uncatalogued co-natal
associations could violate this assumption.

\subsection{Exposing DNA-damage signatures in CRISPR screens}
\label{app:cutting}

\paragraph*{Data and guide selection.}
We use interphase imaging features from the HeLa-TetR-Cas9 screen of Funk
et al.\ \cite{funk2022}. The four primary features are integrated
nuclear DAPI intensity (DNA content), nuclear area, mean nuclear
$\gamma$H2AX intensity and $\gamma$H2AX foci count. Except for raw foci count, features use the authors' transformations and normalization by
the median and median absolute deviation (MAD) of non-targeting cells in
the same well.

For our reanalysis, we selected a cutting-guide group using genomic and
gene-expression information, independently of the imaging measurements.
To reduce the likelihood of disrupting an active gene, we require target-gene
expression below 1 TPM in DepMap 24Q4, 1 nTPM in the Human Protein Atlas and
1 mean CPM in RNA sequencing of the screened cell line (GEO GSE199901).
Agreement among these three sources reduces reliance on any one expression
dataset. The predicted cut must also lie outside expressed genes and their
1\,kb flanks (GENCODE v44), to limit potential effects on overlapping or
neighboring genes. To limit potential cutting at additional sites, each guide
must have a single exact genomic target adjacent to an NGG recognition
sequence (where N denotes any DNA base), with no additional NGG-adjacent
target differing by one base. We require a DepMap absolute copy number of
at least one, indicating that the target locus is present. 

Excluding histone, read-through and duplicated-locus genes leaves 1{,}704
guides targeting 462 genes. Of 250 non-targeting guides, 247 remain after
excluding three with a genomic site one mismatch away.

\paragraph*{Pooled tests and detection power.}
\pqm{} uses the shared configuration in section~\ref{app:hyperparams}: pooled
standardization, ten tessellations and the same sample-size rule, which gives
20 reference points throughout these pooled comparisons. The other
$\chisq$-type tests follow the definitions in the same section.
The guide-median benchmark compares group means of per-guide feature
medians, standardizing feature differences and taking their maximum.
Standardization uses the observed and all permuted assignments together,
treating each alike. This adapts Funk et al.'s aggregation \cite{funk2022},
with whole-guide permutations and no aggregation into genes. As post-hoc
benchmarks, guide means replace medians in the same procedure, and the
foci-fraction test compares guides' fractions of cells with at least one focus.
For descriptive foci fractions, we use all cells per eligible guide and
average the guide-level fractions with equal weights.

Eligible guides have at least 100 cells. For detection curves, genes targeted by cutting guides and non-targeting guides are divided
into disjoint halves (231 genes each; 124 and 123 control guides). Within each
half, $G$ genes (one guide each) and $G$ controls are sampled 20 times at
$G\in\{4,8,16,32,64,120\}$, with 40 cells per guide and 99 whole-guide
label permutations \cite{winkler2014}.
PTEs randomize ties as
$(\#\{T_b>T\}+U[1+\#\{T_b=T\}])/(B+1)$, where $T$ is the observed
statistic, $T_b$ are the $B$ permuted statistics and $U$ is uniform on
$(0,1)$. Each method is assessed separately at $\mathrm{PTE}\le0.05$. Detection
rates pool the two halves; error bars are Wilson intervals.
We estimate the guide count needed for 50\% detection by fitting
nondecreasing detection curves and interpolating between sampled guide
counts on a logarithmic scale. The 95\% intervals use 1{,}000 bootstrap
resamplings of whole replicate curves, keeping the tests paired. All benchmarks use the same 240 subsamples, permutations and PTE
convention.

\paragraph*{Well matching and cell-cycle checks.}
Cells occupy 46 wells on eight plates imaged on two dates. For matching,
we start with one guide from each of 247 randomly selected genes targeted by cutting guides and all 247 non-targeting guides. Selecting guides with coverage of the same 40 wells, with one guide per
gene, leaves 246 genes and 234 controls. Each guide contributes one cell
per well. Calibration uses
999 whole-guide permutations, preserving the matched well composition.

All tests reject after matching, leaving out each plate in turn, and
analyzing each imaging date separately; matched non-targeting halves do not.
Random and high- versus low-abundance halves of non-targeting guides also
do not reject after multiple-testing correction. Matching preserves
\pqm{}'s sensitivity advantage over medians, with no resolved advantage
over guide means.

The guide-averaged foci-positive fraction is higher for cutting guides
in every well and within every DNA-content phase. Phases use boundaries at
normalized integrated DAPI values of $-1.97$, $1.08$ and $6.59$, estimated
from a three-component mixture fitted without group labels. Reweighting each guide to the controls' phase distribution leaves the excess
unchanged. These checks
assess robustness within this screen, rather than independent replication.

\paragraph*{Localization.}
Figure~\ref{fig:realdata}D uses the initial guide groups from the matched
checks, with 100 cells per guide. Twenty reference points are sampled in
proportion to group sizes; remaining cells join the nearest point in the
four standardized features. Regions are binned by each feature's mean,
using deciles from 500 full-data tessellations and merging tied edges.
Each bin reports the pooled ratio of cells carrying cutting guides to control cells. Lines and
bands show the median and 2.5--97.5 percentiles of 400 guide-resampled
curves, each using 100 cells per guide and 100 tessellations. Each
resampling contributes one curve; control curves compare non-targeting
halves. The four curves are descriptive projections of the same
four-dimensional regions, not separate feature tests.

The guide groups also differ in source library, GC content and abundance,
so the mismatch cannot be attributed uniquely to cutting. The cutting-guide group is a biologically motivated comparison, not a validated reference
for gene-specific inference.

\ifscienceformat
\PQMassMethodFigure
\BaselineNoiseFigure
\CoverageFigure
\PermutationRanksFigure
\CalibrationComparisonFigure
\MultivariateComparisonFigure
\CoaddDepthFigure
\EnsembleSizeFigure
\ReferenceSizeFigure
\fi

\clearpage
\end{document}